\documentclass[aps,prl,groupedaddress,twocolumn,showpacs]{revtex4-1}
\usepackage{graphicx}
\usepackage{dcolumn}
\usepackage{bm}
\usepackage{amssymb}
\usepackage{epstopdf}
\usepackage{amsmath}
\usepackage{color}
\begin{document}

\title{Prediction of single-atom-thick transition metal nitride CrN$_4$ with a square-planar network and high-temperature ferromagnetism}
\author{Dapeng Liu$^{1}$}
\author{Panjun Feng$^{1}$}
\author{Shuo Zhang$^{1}$}
\author{Miao Gao$^{2}$}
\author{Fengjie Ma$^{3}$}
\author{Xun-Wang Yan$^{1}$}\email{Corresponding author: yanxunwang@163.com}
\author{Z. Y. Xie$^{4}$}\email{Corresponding author: qingtaoxie@ruc.edu.cn}
\date{\today}
\affiliation{$^{1}$College of Physics and Engineering, Qufu Normal University, Qufu, Shandong 273165, China}
\affiliation{$^{2}$Department of Physics, School of Physical Science and Technology, Ningbo University, Zhejiang 315211, China}
\affiliation{$^{3}$The Center for Advanced Quantum Studies and Department of Physics, Beijing Normal University, Beijing 100875, China}
\affiliation{$^{4}$Department of Physics, Renmin University of China, Beijing 100872, China.}
\begin{abstract}
 Single-atom-thick two-dimensional materials such as graphene usually have a hexagonal lattice while the square-planar lattice is uncommon in the family of two-dimensional materials. Here, we demonstrate that single-atom-thick transition metal nitride CrN$_4$ monolayer is a stable free-standing layer with a square-planar network.
  The stability of square-planar geometry is ascribed to the combination of N=N double bond, Cr-N coordination bond,
 and $\pi$-d conjugation, in which the double $\pi$-d conjugation is rarely reported in previous studies.
 This mechanism is entirely different from that of the reported two-dimensional materials,
 leading to lower formation energy and more robust stability compared to the synthesized g-C$_3$N$_4$ monolayer.
 On the other hand, CrN$_4$ layer has a ferromagnetic ground state, in which the ferromagnetic coupling between two Cr atoms is mediated by electrons of the half-filled large $\pi$ orbitals from $\pi$-d conjugation.
 The high-temperature ferromagnetism in CrN$_4$ monolayer is confirmed by solving the Heisenberg model with Monte Carlo method.

\end{abstract}


\maketitle
Two-dimensional (2D) materials possess a variety of unique electronic properties due to the reduced dimensionality, which have attracted tremendous research interest and become the focused issues of frontier research in physics and material fields.
After graphene was discovered in 2004 \cite{Novoselov2004}, various 2D materials have been fabricated in experiments, such as boron nitride \cite{Song2010}, silicene \cite{Lalmi2010}, borophene \cite{Mannix2015}, stanene \cite{Saxena2016}, transition metal dichalcogenides \cite{Coleman2011}, MXenes \cite{Naguib2011}, $etc$.
In theoretical studies, many kinds of 2D materials were proposed \cite{Ashton2017}.
Among them, only a small fraction of 2D materials consist of single layer of atoms such as graphene, while others consist of several atomic layers such as MoS$_2$.
We call the former single-atom-thick 2D materials, which usually have hexagonal honeycomb geometry and are composed of the third to fifth group elements.
Because no magnetic metal element is included in these single-atom-thick 2D materials, there is no intrinsic magnetism in them and it limits their application in magnetic devices.

Recently, the room-temperature ferromagnetism in CoN$_4$-embedded graphene was realized and the Curie temperature reached up to 400 K, in which 3$d$ metal Co atoms were anchored in the graphene plane by the aid of N atoms \cite{Hu2021}. In fact, more than twenty kinds of metal atoms can be implanted into graphene plane in the form of MN$_4$ (M = metal) moiety in the experiments of single-atom catalysts synthesis \cite{He2019,LAI2020}, and the planar geometry of MN$_4$ in graphene sheets has been definitely observed by the systematic X-ray absorption fine structure analyses and direct transmission electron microscopy imaging \cite{Fei2018}.
In theoretical works, the structural stability of CoN$_4$C$_{10}$, CoN$_4$C$_2$, and CrN$_4$C$_2$ monolayers consisting of MN$_4$ are demonstrated by means of first-principles calculations \cite{Liu2021,Liu2021a,Liu2021b}. The above studies indicate that MN$_4$ moiety is a special structural component.
A natural question is whether there is a free-standing monolayer composed of only MN$_4$ moiety.
On the other hand, another compound catches our attention. Triclinic beryllium tetranitride BeN$_4$ was synthesized under the pressure of 85 GPa very recently \cite{Bykov2021}, and it transforms to layered van der Waals bonded BeN$_4$ with small exfoliation energy. The planar BeN$_4$ layer with single-atomic thickness is regarded as a new class of 2D materials, which is made up of only BeN$_4$ moieties linked together with a staggered alignment. The stability of BeN$_4$ monolayer further provides a powerful evidence for the high feasibility to synthesize the planar MN$_4$ (M = 3$d$ metal) monolayer.

In a recent theoretical study,
MgN$_4$, PtN$_4$, RhN$_4$, and IrN$_4$ layers with a similar structure to BeN$_4$ monolayer are proved to be stable by first-principles calculations,
while the free-standing NiN$_4$ and CuN$_4$ layers are found to be unstable \cite{Mortazavi2021}.
Because Ref. \citenum{Mortazavi2021} focuses on the nonmagnetic analogues of BeN$_4$ layer, other magnetic metals such as Cr have not been considered in their work.
Therefore, we want to clarify two following questions.
One is whether CrN$_4$ moiety can form planar single-atom-thick layers, and the other is that if so, what structure would be formed.

The reason why we choose Cr element is that Cr is a typical transition metal element and majority of chromium compounds are magnetic materials.
What's more, the planar square CrN monolayer with a ratio of Cr:N = 1:1 and its robust ferromagnetism were predicted in the previous study \cite{Zhang2015}.
CALYPSO package \cite{Tang2019} is a famous software for crystal structure searching and has been used extensively.
Using the two-dimensional structure search function in CALYPSO software, we do structure search with the fixed ratio of Cr:N = 1:4 (See Appendix B),
and a square-CrN$_4$ sheet with the lowest energy is screened out. Such a square porous network looks unexpected and has not been reported in previous studies.
In this paper, we will focus on the distinctive square-CrN$_4$ monolayer, demonstrate the stability, analyze the stability mechanism, confirm the high-temperature ferromagnetism, and uncover the origin of ferromagnetism.

The calculations are performed in VASP package, in which the plane wave pseudopotential method and the projector augmented-wave (PAW) pseudopotential with Perdew-Burke-Ernzerhof (PBE) functional \cite{PhysRevB.47.558, PhysRevB.54.11169, PhysRevLett.77.3865, PhysRevB.50.17953} are adopted, and also the nonempirical strongly constrained and appropriately normed (SCAN) meta-GGA method, GGA + U method, the Heyd-Scuseria-Ernzerhof screened hybrid functional (HSE) method are employed to consider the correction of electron correlation \cite{Sun2015,Cococcioni2005,Krukau2006}.
The plane wave basis cutoff is 800 eV and the thresholds are 10$^{-5}$ eV and 0.001 eV/\AA ~ for total energy and force convergence.
The interlayer distance was set to 18 \AA~ and a mesh of $24\times 24\times 1$ k-points is used for the Brillouin zone integration.
The phonon calculations are carried out with the supercell method in the PHONOPY program,
and the real-space force constants of supercells were calculated using density-functional perturbation theory (DFPT) as implemented in VASP  \cite{Togo2015}.
The force convergence criterion (10$^{-5}$ eV/\AA) was used in structural optimization of the primitive cell before building the supercell.
In the ab initio molecular dynamics simulations,
the 3 $\times$ 3 $\times$ 1 supercells were employed and the temperature was kept at 1000 K for 10 ps with a time step of 1 fs in the canonical ensemble (NVT) \cite{Martyna1992}.
The temperature of phase transition in CrN$_4$ system is evaluated by the Monte Carlo method enclosed in the software package developed by Yehui Zhang $et~al.$\cite{Zhang2021}, and the 100 $\times$ 100 $\times$ 1 lattice is used in the Monte Carlo simulation.

 \begin{figure}
\begin{center}
\includegraphics[width=6.0cm]{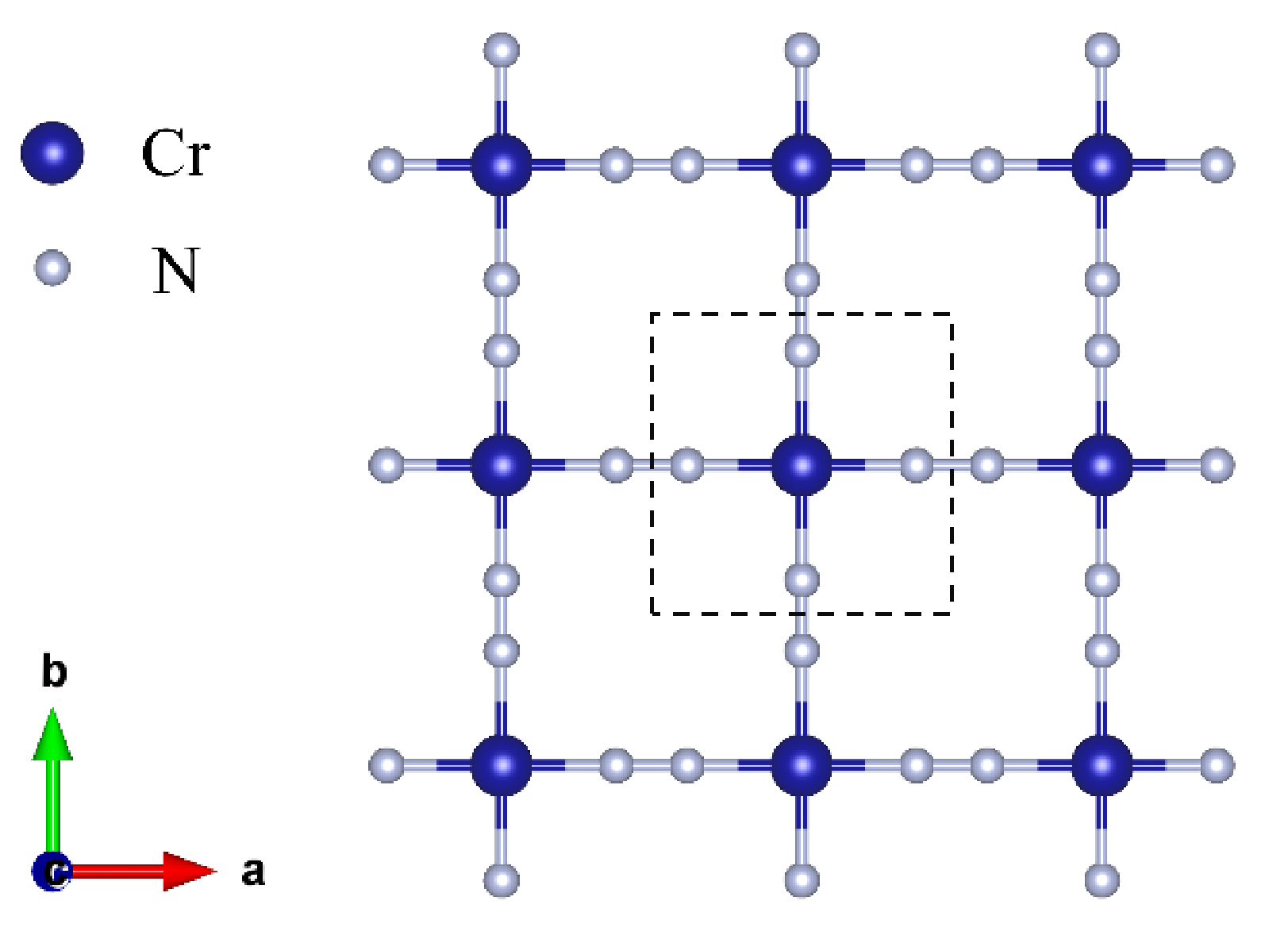}
\caption{Atomic structure of square-CrN$_4$ monolayer with all atoms in a plane. The unit cell is marked with a dashed line square.
 } \label{structmodel}
\end{center}
\end{figure}

The atomic structure of square-CrN$_4$ monolayer is shown in Fig. \ref{structmodel}. The basic structural unit is CrN$_4$ moiety, in which Cr atom is coordinated by four N atoms and located at the center of N atom square. The units are aligned with a top-and-bottom pattern and connected each other by N=N double bonds, forming a porous square network.
The primitive unit cell consists of one Cr atom and four N atoms, and the lattice parameters $a$ = $b$ = 4.92 \AA. The lengths of N-Cr and N=N double bonds are 1.88 \AA~ and 1.16 \AA, respectively.

What mechanism results in the stability of square-planar CrN$_4$ monolayer?
The first thing we notice is the N=N double bond between two adjacent CrN$_4$ units, which is stronger and shorter than N-N single bond because the energy of an N=N double bond is 2.17 times that of an N-N single bond.
Through these N=N double bonds, the CrN$_4$ moieties are tightly and firmly linked together and make up a stable configuration with low total energy.
The next thing to note is the N-Cr coordination bond inside CrN$_4$ moiety, in which N atom provides lone pair electrons and Cr atom provides empty orbitals.
For Cr atom, the 4$s$, 4$p_x$, 4$p_y$, and 3$d_{x^2-y^2}$ orbitals interact to form four $dsp^2$ hybridized orbitals.
This is a common situation in phthalocyanine compounds and some 2D metal-organic frameworks \cite{Sauvage1982,Wang2021}.
Apart from these, the $\pi$ bond of N=N is coupled to the Cr $d_{xz}$, $d_{yz}$, and $d_{xy}$ orbitals to form $\pi-d$ conjugation,
which further increases the strength of Cr-N bond, lowers the total energy, and enhances the robustness of planar structure.

\begin{figure}[htbp]
\begin{center}
\includegraphics[width=8.0cm]{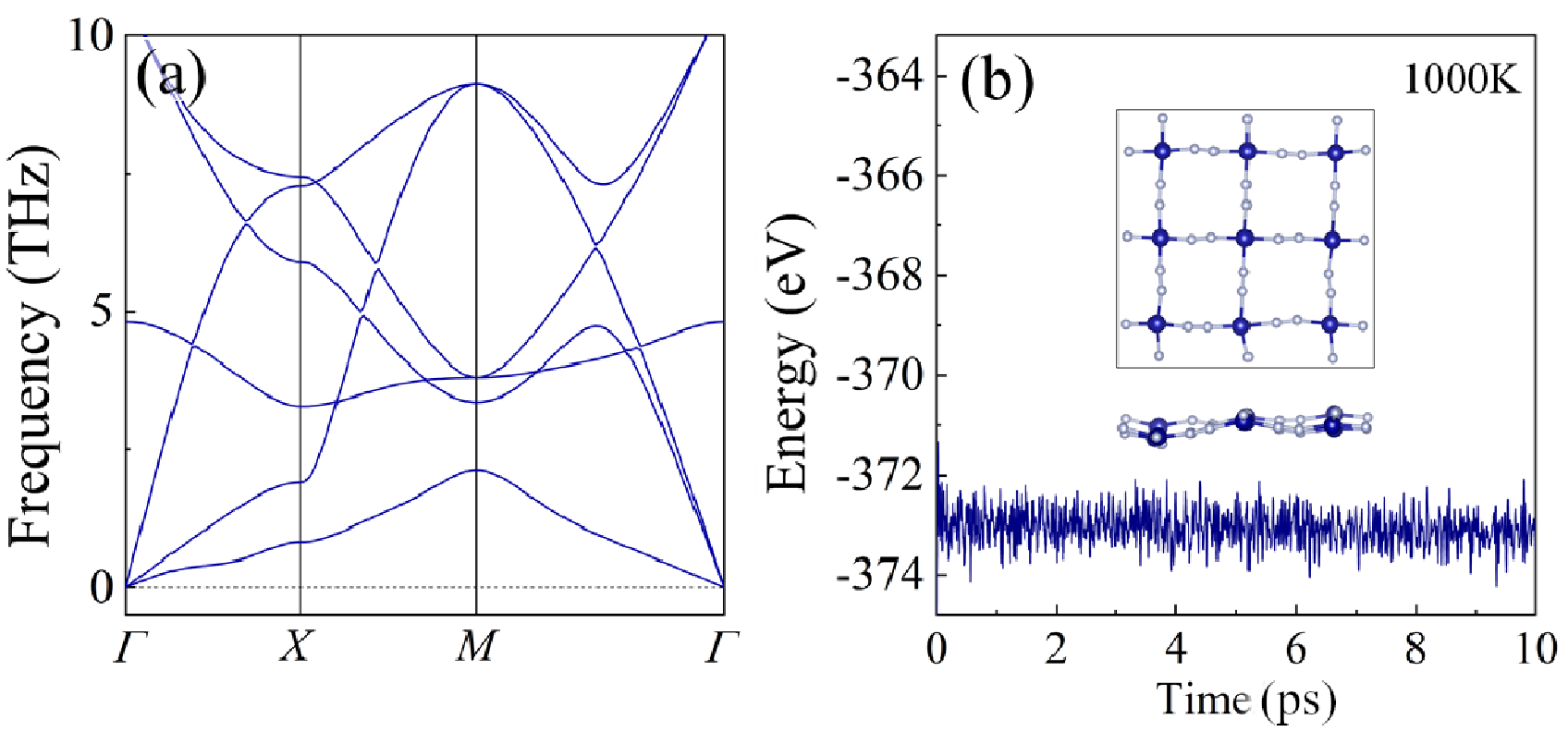}
\caption{(a) Phonon spectra of CrN$_4$ monolayer. (b) The evolution of total potential energy with the time at 1000 K for 10 ps, where the inserts are the top and side views of the final configuration of CrN$_4$ monolayer after 10 ps molecular dynamics simulation.
  } \label{phonon-MD}
\end{center}
\end{figure}

After analyzing the bonding features in CrN$_4$ layer, we then compute the phonon dispersion and do the molecular dynamics simulations to inspect the dynamical and thermal stability.
Fig. \ref{phonon-MD} displays the phonon curves of CrN$_4$ layer. Among them, three acoustic modes start from $\Gamma$ point and no imaginary frequency is observed.
The molecular dynamics simulation of CrN$_4$ layer is performed at the temperature of 1000 K for the time of 10 ps.
We find that total potential energy fluctuates around a certain value and no distinct drop of the energy emerges,
and also the final structure of CrN$_4$ monolayer remains its original framework and no bond is broken.
Therefore, these results demonstrate that the CrN$_4$ monolayer has good thermal and dynamical stability.

\begin{figure}[htbp]
\begin{center}
\includegraphics[width=7.0cm]{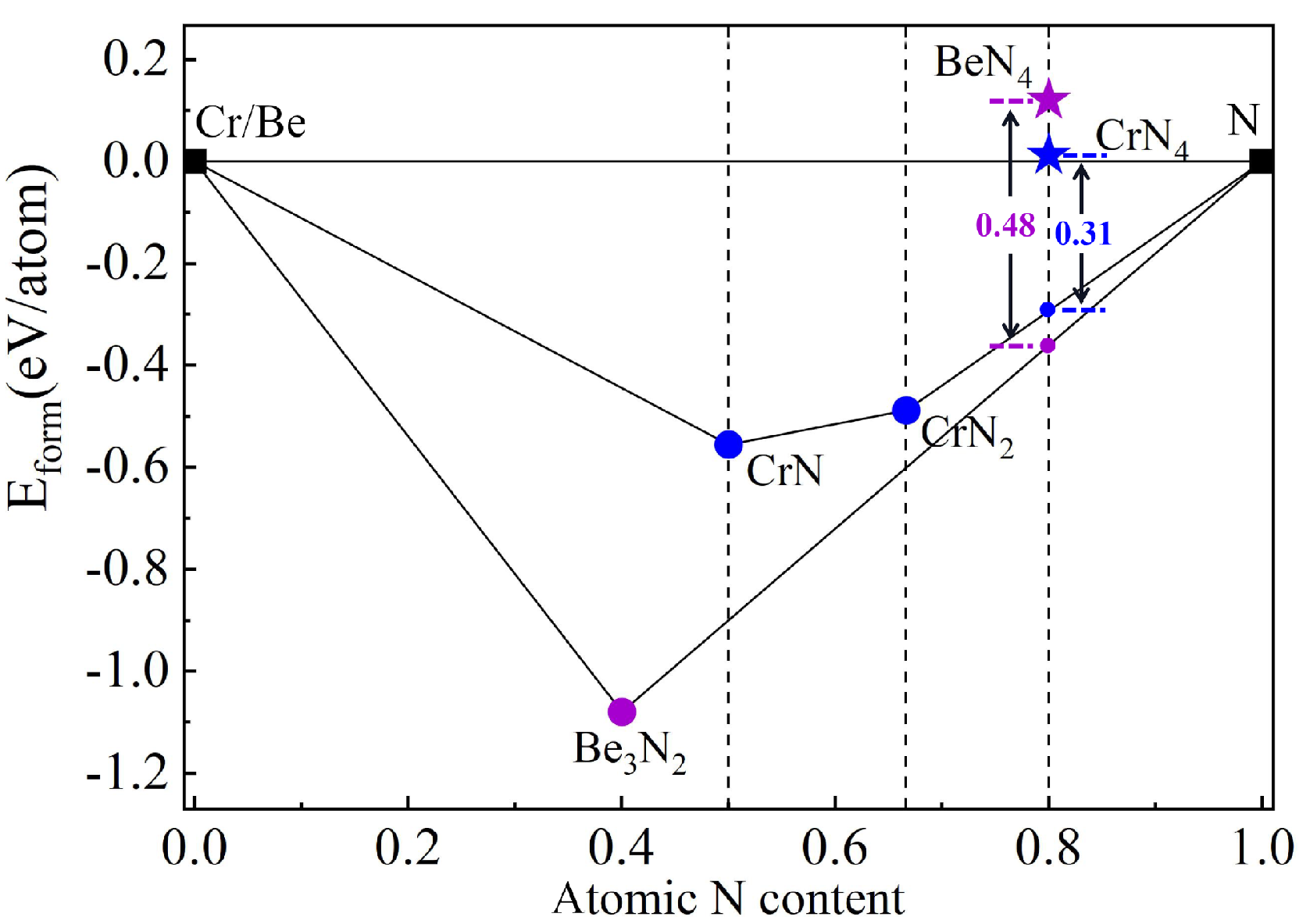}
\caption{ The convex hull diagram of Cr-N and Be-N system.
  } \label{hull}
\end{center}
\end{figure}

The mechanical stability of 2D materials can also be examined by the elastic constants and Young¡¯s modulus \cite{Andrew2012}. We compute the elastic constants of CrN$_4$ sheet and the values of $C_{11}$ = $C_{22}$ = 127.1 N/m, $C_{12}$ = $C_{21}$ = 8.7 N/m, and $C_{66}$ = 2.3 N/m are obtained.
These elastic constants satisfy the two inequalities $C_{11} \cdot C_{22} - C_{12} \cdot C_{21} > 0$ and $C_{66} > 0$, namely, they satisfy the mechanical stability Born criteria \cite{Born:224197}.
According to the formula $E_x$ = ($C_{11} \cdot C_{22}$ - $C_{12} \cdot C_{21}$)/$C_{22}$ and $E_y$ = ($C_{11} \cdot C_{22}$ - $C_{12} \cdot C_{21}$)/C$_{11}$, the in-plane Young¡¯s moduli we obtain are $E_x$ = $E_y$ = 126.5 N/m.
The Young's moduli of CrN$_4$ monolayer are comparable to the ones of SiC, GeC, and BeC monolayers, which are 163.5 N/m, 140.1N/m, and 145.54 N/m, respectively \cite{Andrew2012,Liu2017}.
Hence, the CoN$_4$ monolayer is mechanically stable.
The formation energy of CrN$_4$ monolayer is computed in terms of the expression
$E_{form} = \frac{1}{n}*(E_{tot} - E_{metal} - 2 E_{N_2})$,
 in which $E_{tot}$, $E_{metal}$, and $E_{N_2}$ are the total energy, bulk metal energy per atom, and nitrogen molecule energy, respectively.
For comparison, we also compute the formation energies of $g$-C$_3$N$_4$ and BeN$_4$ monolayers synthesized in experiments and they are 0.346 eV and 0.121 eV.
The calculated formation energy of CrN$_4$ monolayer is 0.013 eV, much smaller than the ones of $g$-C$_3$N$_4$ and BeN$_4$ monolayers.
Fig. \ref{hull} shows the binary phase diagrams of Cr-N and Be-N compounds, in which bulk CrN\cite{Ettmayer1978},
CrN$_2$\cite{Niwa2019}, Be$_3$N$_2$\cite{Reckeweg2003}, metal Cr\cite{BRADLEY1926}, metal Be\cite{Martin1959}, and gas N$_2$ make up the convex hull.
The hull energy of CrN$_4$ monolayer is 0.31 eV, less than the hull energy 0.48 eV of BeN$_4$ monolayer.
The energies of BeN$_4$ and CrN$_4$ above zero in Fig. \ref{hull} are just their formation energies of 0.121 eV and 0.013 eV.
The formation energy and hull energy of CrN$_4$ are comparable to the ones of $g$-C$_3$N$_4$ and BeN$_4$ monolayers synthesized in experiments, 
indicating that it is highly feasible to fabricate the CrN$_4$ monolayer by the high-pressure synthesis method, similar to the synthesis of BeN$_4$ compound \cite{Bykov2021}.

\begin{figure}[htbp]
\begin{center}
\includegraphics[width=8.5cm]{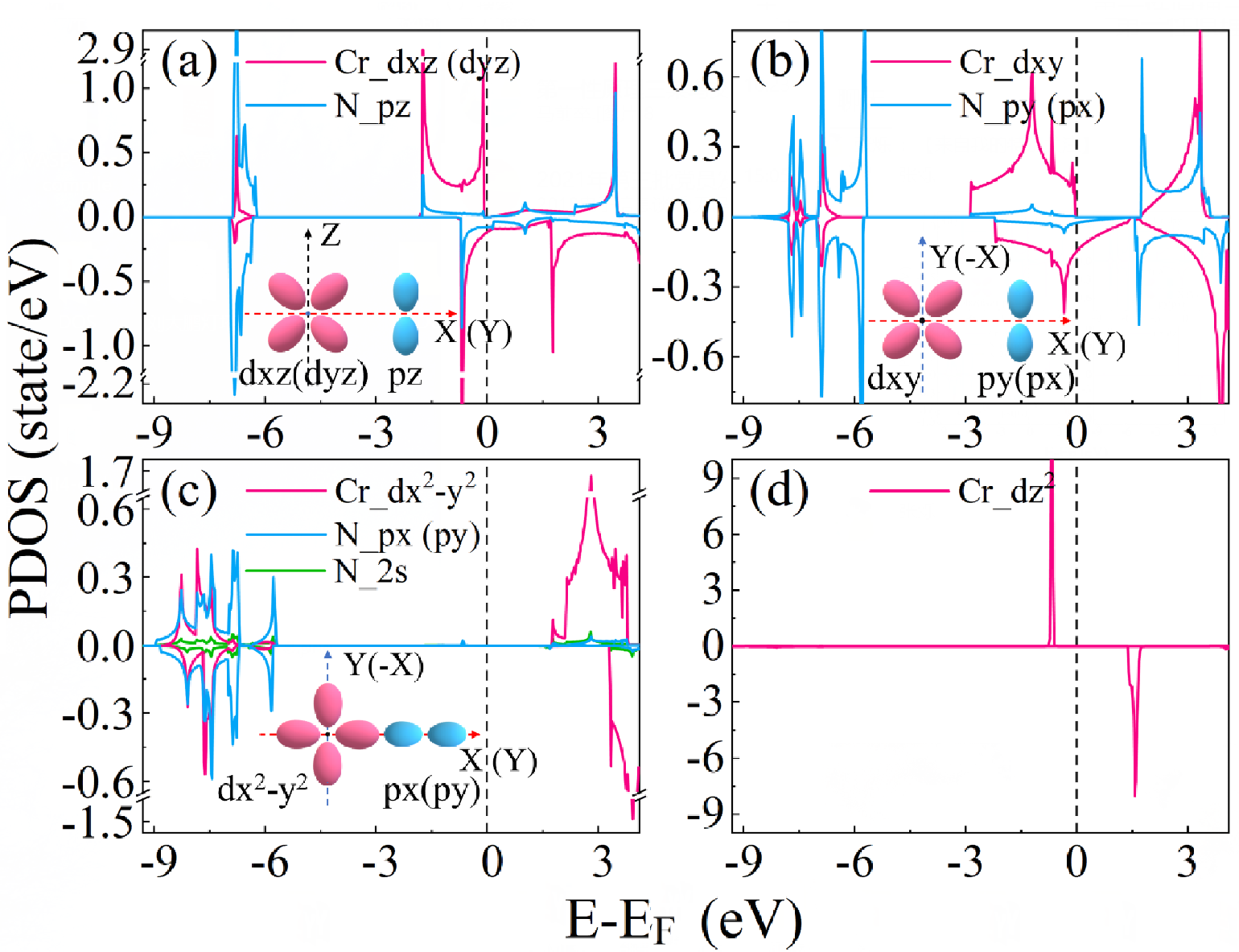}
\caption{Partial density of states of Cr 3$d$ and N 2$p$ suborbitals in CrN$_4$ sheet. The insets are the diagrams of the relevant orbital shape.
  } \label{dos-Cr}
\end{center}
\end{figure}

\begin{figure}[htbp]
\begin{center}
\includegraphics[width=8.2cm]{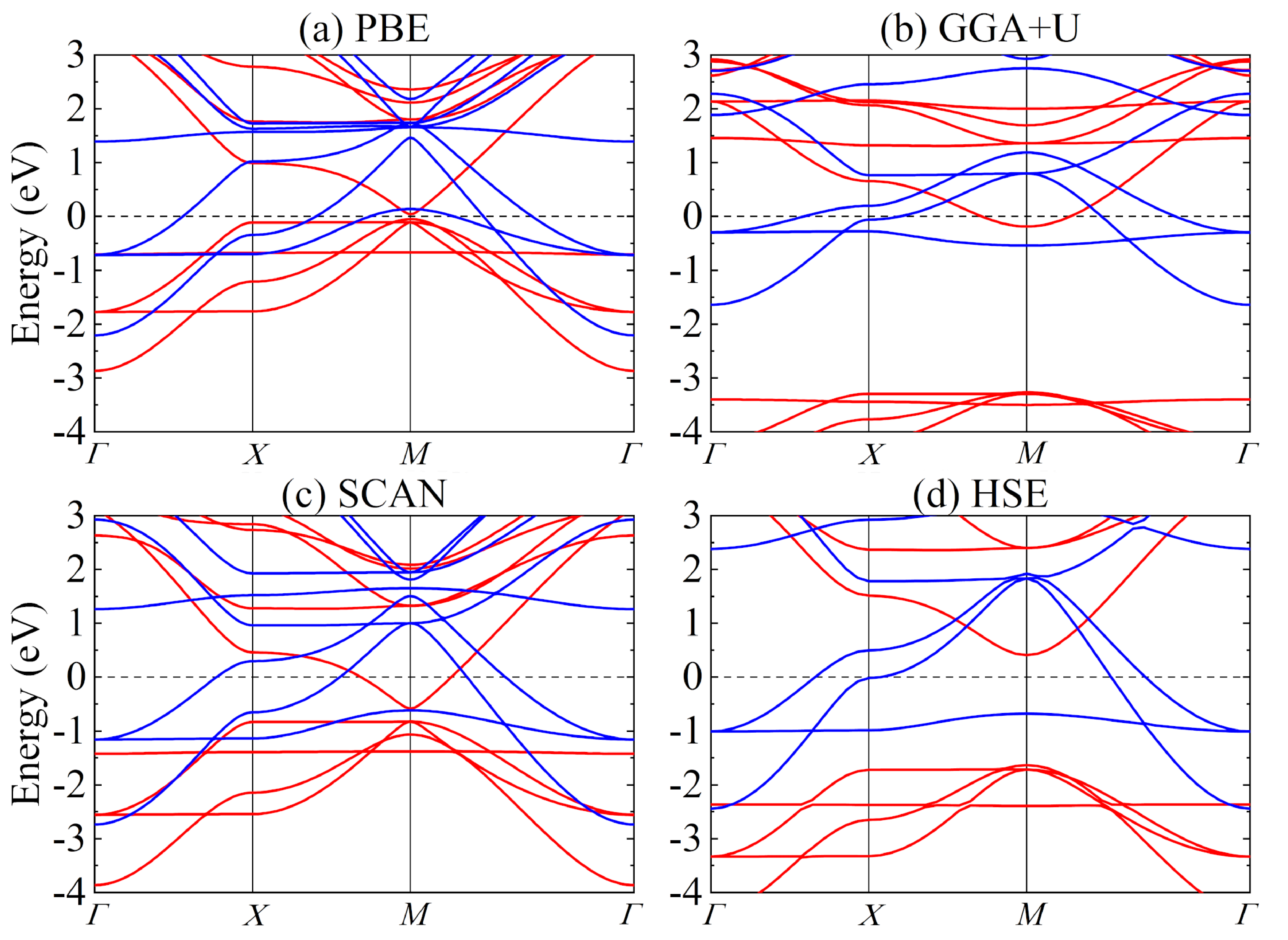}
\caption{Band structure of CrN$_4$ monolayer in ferromagnetic phase with different functional methods, (a) PBE, (b) GGA + U, (c) SCAN, (d) HSE.
The Fermi energy is set to zero.
  } \label{band}
\end{center}
\end{figure}

Then, we investigate the electronic structure of CrN$_4$ layer.
Cr atom is located at a square plane field, whose plane symmetry is P4m affiliated to point group D$_4$. The five partial $d$ orbitals are divided into four group, namely d$_{xz}$/d$_{yz}$, d$_{z^2}$, d$_{xy}$, and d$_{x^2-y^2}$.
In Fig. \ref{dos-Cr}(a), the partial density of states of Cr d$_{xz}$ (d$_{yz}$) and N p$_z$ orbitals are displayed, and they are distributed in the same energy range and take part in the formation of $\pi$-d conjugated electronic states.
Meanwhile, along $x$ ($y$) axis direction Cr $d_{xy}$ orbital and N $p_y$ ($p_x$) make up another $\pi-d$ conjugation, shown in Fig. \ref{dos-Cr}(b).
Namely, along Cr-N=N-Cr chain in $x$ ($y$) direction there are two kinds of $\pi-d$ conjugation, one is related to $d_{xz}$-$p_z$-$p_z$-$d_{xz}$ ($d_{yz}$-$p_z$-$p_z$-$d_{yz}$) orbitals and the other is $d_{xy}$-$p_y$-$p_y$-$d_{xy}$ ($d_{xy}$-$p_x$-$p_x$-$d_{xy}$) orbitals.
It is probably the first time that the double $\pi$-d conjugation situation is reported.
Fig. \ref{dos-Cr}(c) displays the density of states of Cr $d_{x^2-y^2}$, N $p_x$ ($p_y$) and $s$ orbitals. The three orbitals are related to the Cr-N coordination bond for which the $d_{x^2-y^2}$ orbital is empty and lone pair electron comes from the hybridized states of N $p_x$ ($p_y$) and $s$ orbitals.
The density of states of Cr $d_{z^2}$ orbitals are presented in Fig. \ref{dos-Cr}(d), and its spin-up states are fully occupied and the spin-down states are empty. The spin polarization of electrons in $d_{z^2}$, $d_{xz}$, $d_{yz}$, and $d_{xy}$ orbitals gives rise to the magnetic moment of 2 $\mu_B$ around Cr atom.
Fig. \ref{band}(a)-(d) display the energy band structures of CrN$_4$ monolayer in ferromagnetic phase with PBE, GGA + U, SCAN, and HSE functional methods, respectively.
 The red and blue curves correspond to the spin-up and spin-down bands. In the energy range from -4 to 0 eV,
 the positions of red curves change obviously,
 which reflects the considerable influence of different functionals on electronic structure of CrN$_4$. There are two hole-type bands (blue) and one electron-type band (red) crossing the Fermi energy in Fig. \ref{band}(b) and (c), while only two blue curves cross the Fermi energy in Fig. \ref{band}(d), indicating the larger exchange splitting in HSE calculations than in GGA +U and SCAN calculations. The metallic properties of CrN$_4$ monolayer are tightly associated with these bands through the Fermi energy, which are made up of the delocalized electronic states due to the double $\pi$-d conjugation mentioned above. 

Flat band is an unusual characteristic of Bloch electronic states in condensed matter physics. Due to weak dispersiveness, there are the small band width and high density of states, resulting in the small kinetic energy and large Coulomb potential energy. Because the Coulomb interaction is far greater than the kinetic energy of electrons in the flat band states, the associated compounds exhibit some exotic strong correlation phenomena, such as superconductivity \cite{Miyahara2007}, ferromagnetism \cite{Zhang2010}, Wigner crystal \cite{Wu2007}, and fractional quantum Hall effect \cite{Neupert2011}.
A sharp peak of the density of states spectra in Fig. \ref{dos-Cr}(d) is associated with the flat band around -0.7 eV below the Fermi energy in Fig. \ref{band}(a). Due to the 2D structure, there is no atom above or below the CrN4 plane. In this special 2D crystal field, the Cr $d_{z^2}$ orbital does not hybridize with the orbital from other atoms and keeps an isolated state, which results in the sharp peaks of density of states and the flat bands. The flat bands at -3.4 eV in Fig. \ref{band}(b), -1.4 eV in Fig. \ref{band}(c), and -2.4 eV in Fig. \ref{band}(d) all come from the dz2 orbitals in the GGA+U, SCAN, and HSE functionals calculations, which are closely related to the local part of Cr moment.

\begin{figure}[htbp]
\begin{center}
\includegraphics[width=7.0cm]{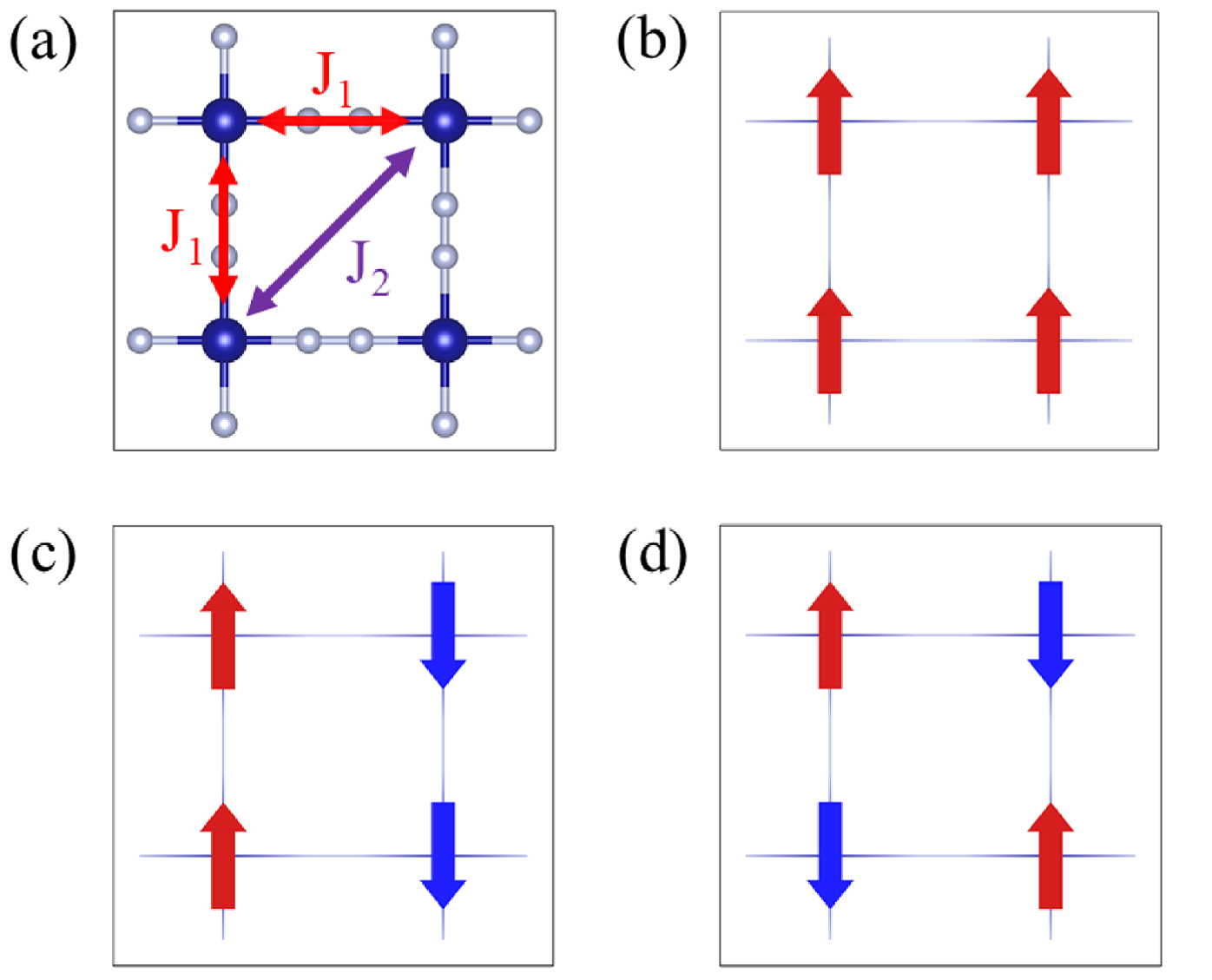}
\caption{(a) Top view of atomic structure of CrN$_4$. $J_1$ and $J_2$ are the nearest and next-nearest neighbored exchange interaction. (b)$\sim$(d) The sketches of FM, coll-AFM, and neel-AFM.
  } \label{orders}
\end{center}
\end{figure}

\begin{table*}
	\caption{The energies in FM, coll-AFM, and neel-AFM orders, local magnetic moment($Moment$), exchange coupling $J_1$ and $J_2$, the estimated Curie temperatures ($T_{c0}$ and $T_{c1}$) in terms of $J_1 \cdot 2/ln(1+\sqrt{2})$ and Olsen $et~al$'s expression, and the calculated Curie temperature ($T_{c}$) by Monte Carlo method. The units of energy, magnetic moment, couplings $J_1$ and $J_2$, magnetic anisotropy energy, and Curie temperatures are meV, $\mu_B$, meV/$S^2$, meV/$S^2$, and K.
	}
	\label{energy-mom}
	\renewcommand\tabcolsep{6.5pt} 
\begin{tabular*}{13.5cm}{llllccccccc}
		\hline
		Method & $E_{\mathit{FM}}$ & $E_{\mathit{coll}}$ & E$_{\mathit{neel}}$  & $Moment$ & $J_1$  & $J_2$ & A &$T_{c0}$ &$T_{c1}$ &$T_{c}$ \\
		\hline
		PBE  &0.0  & 34.4   & 46.8  &2.0  & -11.7 & -2.7 & -0.303 &309 &47 &121  \\
		GGA+U &0.0 & 115.0 & 124.7  &2.1 & -31.2 & -13.2 & -0.164 &822 &91 &356 \\
		SCAN &0.0  & 88.8   & 143.4 &2.0 & -35.8 & -4.3  & -0.215 &944 &94 &300 \\
		HSE &0.0  & 197.8   & 209.8 &2.0 & -52.6 & -23.2 & -0.291 &1382&144 &620 \\
		\hline
\end{tabular*}
\end{table*}
Ferromagnetic 2D materials have many potential applications in the next-generation spintronic devices, but low Curie temperature is the main obstacle to hinder the applications. Therefore, exploring 2D ferromagnetic materials with high critical temperature is a significant topic in physics and material fields.
Next, we demonstrate that CrN$_4$ monolayer is a high-temperature ferromagnetic monolayer.
To determine the magnetic ground state of CrN$_4$ monolayer, we perform the spin-polarized calculations for several magnetic orders, including ferromagnetic order (FM), collinear antiferromagnetic order (coll-AFM), and Neel antiferromagnetic order (neel-AFM). The atomic structure and three magnetic orders are sketched in Fig. \ref{orders}, and for clarity, the atomic structure is displayed with the wire frame in Fig. \ref{orders}(b), (c), and (d).
To confirm the ferromagnetic ground state of CrN$_4$, the PBE, GGA+U, SCAN, and HSE functional methods are used separately to calculate the electronic structures of CrN$_4$ monolayer, and the total energies per formula cell and magnetic moments around Cr atoms are listed in Table. \ref{energy-mom}.
The value of Hubbard U is 5.46 eV, which is derived from the self-consistent calculation with linear response method \cite{Cococcioni2005}.
As can be seen, the energy values from different functional methods for FM, coll-AFM, and neel-AFM states have the same order, which strongly demonstrates that the single-atom-thick CrN$_4$ layer is a ferromagnetic 2D material.
The ferromagnetic coupling between two Cr moments is mediated by N=N double bond.
The physical picture is that on the account of Hund's rule, the electron in $d_{z^2}$ and $d_{xz}$ orbital of each Cr atom has the same spin.
Along a Cr-N=N-Cr chain in $x$ axis direction, the $d_{xz}$, $p_z$, $p_z$, and $d_{xz}$ are recombined to construct the delocalized $\pi$ states and they are half-filled. So, the $d$ electrons in these delocalized $\pi$ states have the same spin because of Hund's rule.
Consequently, $d$ electrons in $d_{xz}$ and $d_{z^2}$ orbitals belonging to two Cr atoms have the same spin and their magnetic coupling is ferromagnetic.
We can draw the same conclusion when the magnetic coupling is analyzed in terms of $d_{yz}$ or $d_{xy}$ orbitals.

The magnetic moment of Cr atom in CrN$_4$ originates from the partially occupied $d$ orbitals, shown in Fig. \ref{dos-Cr}. The spin-up channel of $d_{z^2}$ orbital is fully occupied and its spin-down channel is empty, which is related to the local moment of 1.0 $\mu_B$ of Cr atom. The $d_{xy}$, $d_{yz}$, and $d_{xz}$  orbitals distribute around Fermi energy and make a main contribution to the metallic behavior of CrN$_4$ monolayer. Each of $d_{xy}$, $d_{yz}$, and $d_{xz}$ orbitals provides the moment of 0.33 $\mu_B$, which corresponds to the itinerant part of Cr moment. So, there are both localized moment and itinerant moment in CrN$_4$ monolayer. Just as the Heisenberg model is used for FeSe and FeTe compounds \cite{Ma2009,Yu2015}, we use the Heisenberg model to describe the magnetic interactions in CrN4 layer.
The Hamiltonian is defined as

\begin{equation}
\label{Heisenberg}
H = J_{1}\sum_{<ij>}\vec{{S}_{i}}\cdot \vec{{S}_{j}} + J_{2}\sum_{\ll ij^{\prime} \gg}\vec{{S}_{i}}\cdot \vec{{S}_{j^{\prime}}} + A\sum_{i}(S_{iz})^2,
\end{equation}
where $j$ and $j^\prime$ denote the nearest and next-nearest neighbors of $i$ site.
$J_1$ and $J_2$ are the nearest and next-nearest neighbored couplings, which can be derived from the energy differences among FM, coll-AFM, and neel-AFM orders \cite{Ma2008}.
A is the single-site magnetic anisotropic energy, which is the energy difference when the Cr moment is along (1 0 0) and (0 0 1) directions.
The relevant data are shown in Table. \ref{energy-mom}.
For the CrN$_4$ monolayer, the next-next-neighbor coupling $J_3$ is almost zero according to the SCAN calculations. So, the $J_3$ and other couplings over a longer distance are not contained in the Heisenberg Hamiltonian in Equation \ref{Heisenberg}. The computational details concerning $J_3$ are attached in Appendix A.

\begin{figure}[htbp]
\begin{center}
\includegraphics[width=8.5cm]{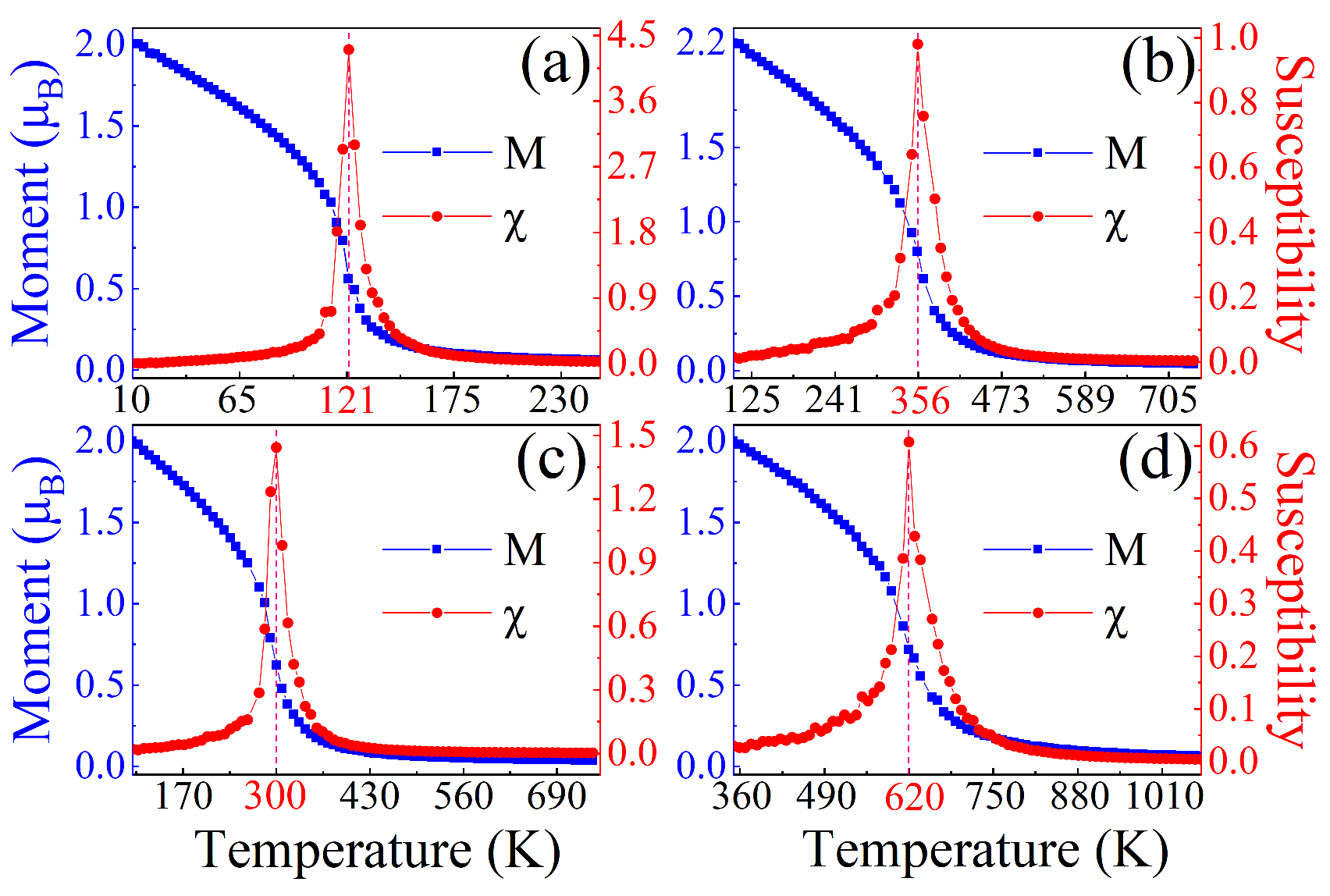}
\caption{The susceptibility $\chi$ and average magnetic moment $M$ as functions of temperature for the Heisenberg model on a square lattice, and the parameters $J_1$, $J_2$ and $A$ of Heisenberg model derived from the various functional calculations, (a) PBE, (b) GGA + U, (c) SCAN, (d) HSE.
  } \label{Tc}
\end{center}
\end{figure}

Curie temperature is a key parameter to determine the practical value of the magnetic materials.
Firstly, we make a rough estimate in terms of the formula $T_c = J_1 \cdot 2/ln(1+\sqrt{2})$ \cite{Onsager1944,Yang2018,Xie2012}, the analytical solution of the Ising model for a two-dimensional square lattice,
and the calculated Curie temperature ($T_{c0}$) are listed in Table. \ref{energy-mom}.
Because Ising model is the limit of Heisenberg model with magnetic anisotropy going to infinity and the magnetic anisotropy is small in most two-dimensional materials, the critical temperature is usually overestimated.
For a magnetic two-dimensional lattice, Heisenberg model is a more suitable model to determine the critical temperature, in which magnetic anisotropy is essential in terms of Mermin-Wagner theorem.
Olsen $et~al.$ proposed an analytical expression on the critical temperature by fitting the results of Monte Carlo simulations \cite{Torelli2019}, $T_c = T_{c0} \cdot tanh^{1/4}[\frac{6}{N_{nn}}log(1+\frac{0.033A}{J})]$, where $N_{nn}$ is the number of nearest neighbors, $T_{c0}$ is the critical temperatures for the corresponding Ising model, A and J are the single-ion magnetic anisotropy and the nearest neighboring exchange coupling, respectively. The critical temperatures obtained from Olsen $et~al$'s expression are listed as $T_{c1}$ in Table. \ref{energy-mom}, which are underestimated because only the nearest exchange coupling J$_1$ is included.
Yehui zhang $et~al.$ developed a software package to compute the critical temperature of two-dimensional magnetic lattice on the base of Heisenberg model and Monte Carlo method, which is greatly successful in estimating the Curie temperature of ferromagnetic CrI$_3$ monolayer\cite{Zhang2021}.
By means of Zhang's software package, we solve the Heisenberg model with the parameters $J_1$, $J_2$ and $A$ in Table \ref{energy-mom}.
The variations of magnetization ($M$) and susceptibility ($\chi$ = $\frac{<\vec{M}^2> - <\vec{M}>^2}{k_BT}$) with respect to temperature are presented in Fig. \ref{Tc} and the Curie temperatures ($T_{c}$) of CrN$_4$ monolayer are 121 K, 356 K, 300 K, and 620 K according to the PBE, GGA + U, SCAN, and HSE functional methods, respectively.

The Curie temperature T$_c$ from the HSE hybrid functionals calculations is obviously higher than the ones from PBE, GGA + U, and SCAN calculations. The reason is that the HSE hybrid functional method tends to overestimate the exchange splitting of $d$ electronic states and result in an increase of the spin-polarization energy due to the introduction of a fixed portion of Fock exchange \cite{Paier2006}. On the other hand, because the electronic correlation effect is not fully considered in the PBE functional, the T$_c$ according to the standard PBE calculations is usually underestimated.
In the GGA +U and SCAN calculations, the electron correlation effect is more reasonably taken into account, which leads to the more credible Curie temperatures.
This is verified by the fact that the T$_c$s of 356 K and 300 K from the GGA +U and SCAN calculations are roughly consistent and can confirm each other.
At present, the reported two-dimensional ferromagnetic materials usually have a low Curie temperature.
As displayed in the recent review paper\cite{Guo2020}, among 44 kinds of ferromagnetic 2D materials the highest Curie temperature predicted on the basis of the Heisenberg model is 261 K. So, the 356 K and 300 K are the high Curie temperature for a two-dimensional ferromagnetic material.
As for the mechanism of high-temperature ferromagnetism, it is closely associated with the special Cr-N=N-Cr chain of the CrN$_4$ monolayer, along which the double $\pi$-d conjugation can give rise to the strong exchange coupling between two neighboring Cr atoms, leading to robust ferromagnetism.
The robust ferromagnetism and specific structure make the single-atom-thick CrN$_4$ monolayer unusual and further study is expected.

In summary, we propose a single-atom-thick two-dimensional compound CrN$_4$ from the first-principles calculations, which is a transition metal nitride sheet with a square-planar network structure rarely reported before. The stability has been verified by the calculations involved in phonon spectra, molecular dynamics simulation, elastic constant, and formation energy, and the mechanism is ascribed to the cooperation of N=N double bond, Cr-N coordination bond, and $\pi$-d conjugation effect.
Especially, the double $\pi$-d conjugation effect is discovered for the first time.
The feasibility of fabrication in experiments is explained by the very small formation energy and the $M$N$_4$ (M = metal) structure units ever synthesized.
More importantly, the CrN$_4$ monolayer is a ferromagnetic 2D compound with the high Curie temperature.

We sincerely thank Prof. Jinlan Wang, Dr. Yehui Zhang, and Prof. Qiang Li in Southeast University in China for sharing their code and very helpful discussions on MC calculations. This work was supported by the National Natural Science Foundation of China (Grants Nos. 11974207, 11974194, 11774420, 12074040), the National R\&D Program of China (Grants Nos. 2016YFA0300503, 2017YFA0302900), and the Major Basic Program of Natural Science Foundation of Shandong Province (Grant No. ZR2021ZD01).

\appendix
 \renewcommand{\appendixname}{Appendix~\Alph{section}}
 \setcounter{table}{0}
\renewcommand{\thetable}{A\arabic{table}}
\setcounter{figure}{0}
\renewcommand{\thefigure}{A\arabic{figure}}
\section{Appendix A: Next-Next neighbor exchange coupling $J_3$}
\begin{figure}[h]
\begin{center}
\includegraphics[width=6.0cm]{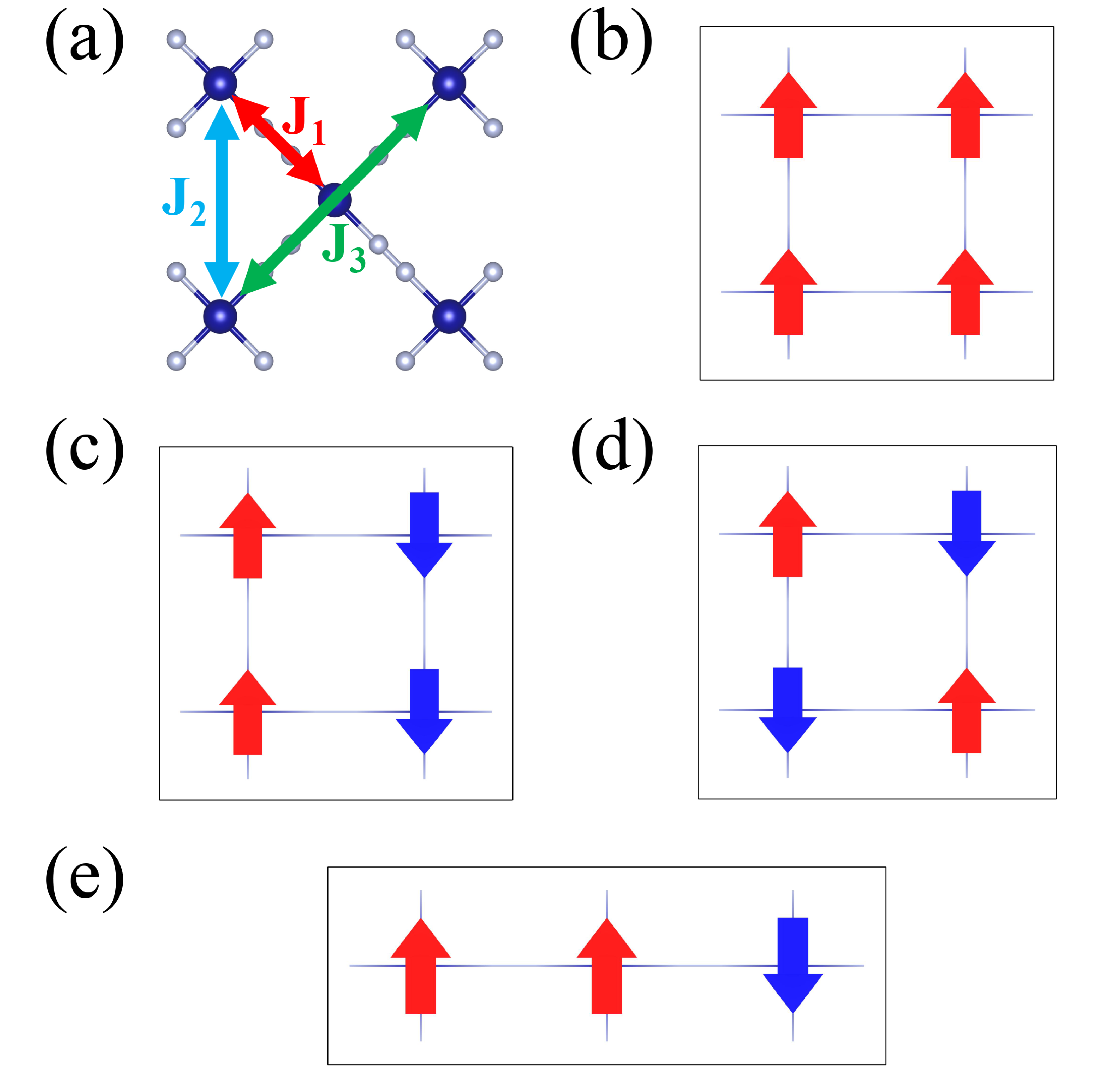}
\caption{(a)Atomic structure of CrN$_4$ monolayer with the exchange coupling $J_1$, $J_2$, and $J_3$. (b) Ferromagnetic order (FM). (c) Antiferromagnetic order I (AFMI). (d) Antiferromagnetic order II (AFMII). (e) Antiferromagnetic order III (AFMIII).
  } \label{figA1}
\end{center}
\end{figure}

We compute the next-next-neighbor exchange coupling $J_3$ with GGA+U and SCAN methods and listed in Table. \ref{J3} (the results from GGA+U and SCAN calculations are more credible, which has been discussed in the last paragraph on Page 5). As can be seen, the $J_3$ from SCAN calculations almost decays to zero. And the $J_3$ from GGA+U calculations also is less than $J_2$, showing an obvious decay. For the GGA+U results, when $J_3$ is included in the Heisenberg Hamiltonian, the Curie temperature must be higher than $Tc$ = 356 K presented in Table. \ref{energy-mom}. Therefore, the Heisenberg Hamiltonian with $J_1$ and $J_2$ only being considered is reliable. For computing the next-next-neighbor exchange coupling $J_3$, the four magnetic orders are adopted and shown in Fig. \ref{figA1}.
\begin{table}[h]
	\caption{The energies per CrN$_4$ formula cell in various magnetic orders and the exchange couplings $J_1$, $J_2$, and $J_3$. The units of energy and exchange coupling are meV and meV/$S^2$. The energy in FM order is set to zero.
	}
	\label{J3}
\begin{tabular*}{8.5cm}{llllcccc}
		\hline
		Method & $E_{\mathit{FM}}$ & $E_{AFMI}$ & E$_{AFMII}$  &E$_{AFMIII}$  & $J_1$  & $J_2$ & $J_3$  \\
		\hline
		GGA+U &0.0 & 114.97 &124.74 & 88.25 & -31.2 & -13.2 & -11.6 \\
		SCAN  &0.0 & 88.82  &143.40 & 59.30 & -35.8 & -4.3  & -0.08 \\
		\hline
\end{tabular*}
\end{table}

\section{Appendix B: Structural Search for CrN$_4$ }
\begin{figure}[h]
\begin{center}
\includegraphics[width=6.0cm]{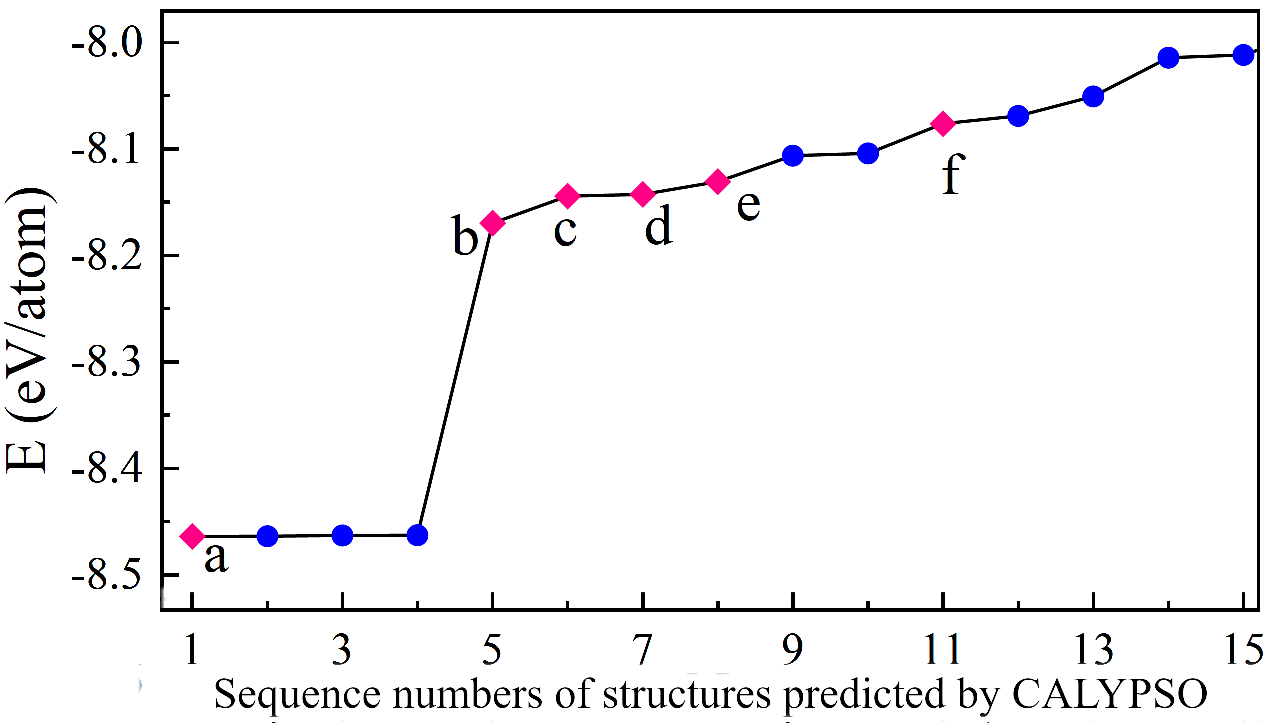}
\caption{Sequence numbers of structures predicted by CALYPSO software. The Red symbols represent the typical structures and the blue symbols mean the repetitive one to the structure represented by the left red symbol.
  } \label{figA2}
\end{center}
\end{figure}

\begin{figure}[h]
\begin{center}
\includegraphics[width=8.5cm]{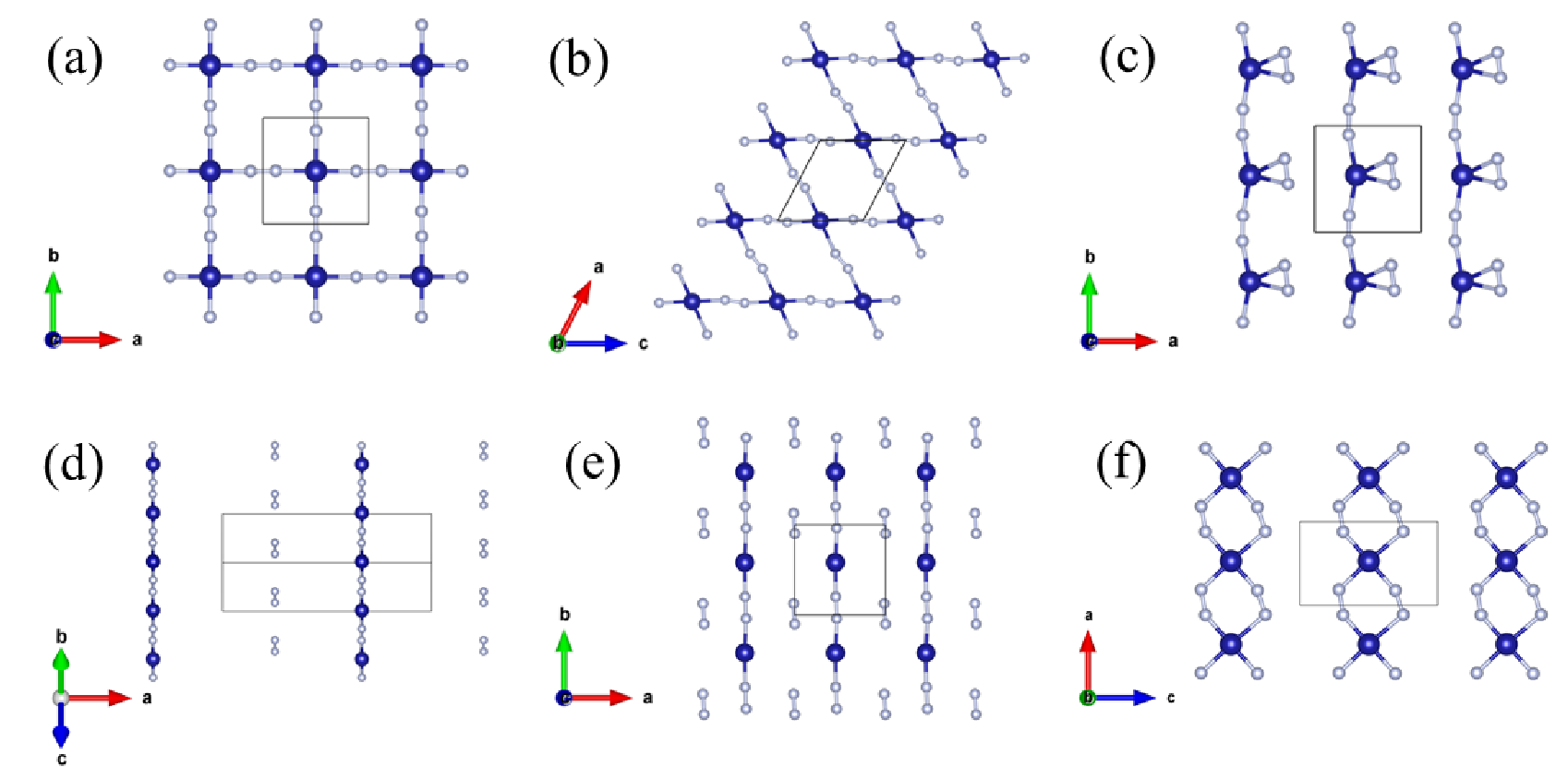}
\caption{The typical structures predicted by CALYPSO software, which energies are displayed in Fig. \ref{figA2} with red symbols.
  } \label{figA3}
\end{center}
\end{figure}

 CALYPSO code is employed to confirm the square structure of CrN$_4$ monolayer. Using two-dimensional structure search function in CALYPSO software, we do structure search with the fixed ratio of Cr:N = 1:4. In Fig. \ref{figA2}, the horizontal axis is the sequence numbers of structures predicted by CALYPSO software and the vertical axis is the energies of different structures. The Red symbols represent the typical structures and the blue symbols mean the repetitive one to the structure represented by the left red symbol. Fig. \ref{figA3} displays the typical structures predicted which are marked with ¡°a, b, c, d, e, f¡± in Fig. \ref{figA2}. The results demonstrate the CrN4 monolayer in our manuscript is the lowest energy structure. Except ¡°f¡± structure, these structures contain the Cr-N=N-Cr chain, which indicates that the Cr-N=N-Cr chain is a low-energy structural unit. It also explains why the CrN4 monolayer composed of Cr-N=N-Cr chains has the lowest energy.

\bibliography{Ref}

\begin{thebibliography}{52}%
\makeatletter
\providecommand \@ifxundefined [1]{%
 \@ifx{#1\undefined}
}%
\providecommand \@ifnum [1]{%
 \ifnum #1\expandafter \@firstoftwo
 \else \expandafter \@secondoftwo
 \fi
}%
\providecommand \@ifx [1]{%
 \ifx #1\expandafter \@firstoftwo
 \else \expandafter \@secondoftwo
 \fi
}%
\providecommand \natexlab [1]{#1}%
\providecommand \enquote  [1]{``#1''}%
\providecommand \bibnamefont  [1]{#1}%
\providecommand \bibfnamefont [1]{#1}%
\providecommand \citenamefont [1]{#1}%
\providecommand \href@noop [0]{\@secondoftwo}%
\providecommand \href [0]{\begingroup \@sanitize@url \@href}%
\providecommand \@href[1]{\@@startlink{#1}\@@href}%
\providecommand \@@href[1]{\endgroup#1\@@endlink}%
\providecommand \@sanitize@url [0]{\catcode `\\12\catcode `\$12\catcode
  `\&12\catcode `\#12\catcode `\^12\catcode `\_12\catcode `\%12\relax}%
\providecommand \@@startlink[1]{}%
\providecommand \@@endlink[0]{}%
\providecommand \url  [0]{\begingroup\@sanitize@url \@url }%
\providecommand \@url [1]{\endgroup\@href {#1}{\urlprefix }}%
\providecommand \urlprefix  [0]{URL }%
\providecommand \Eprint [0]{\href }%
\providecommand \doibase [0]{http://dx.doi.org/}%
\providecommand \selectlanguage [0]{\@gobble}%
\providecommand \bibinfo  [0]{\@secondoftwo}%
\providecommand \bibfield  [0]{\@secondoftwo}%
\providecommand \translation [1]{[#1]}%
\providecommand \BibitemOpen [0]{}%
\providecommand \bibitemStop [0]{}%
\providecommand \bibitemNoStop [0]{.\EOS\space}%
\providecommand \EOS [0]{\spacefactor3000\relax}%
\providecommand \BibitemShut  [1]{\csname bibitem#1\endcsname}%
\let\auto@bib@innerbib\@empty
\bibitem [{\citenamefont {Novoselov}\ \emph {et~al.}(2004)\citenamefont
  {Novoselov}, \citenamefont {Geim}, \citenamefont {Morozov}, \citenamefont
  {Jiang}, \citenamefont {Zhang}, \citenamefont {Dubonos}, \citenamefont
  {Grigorieva},\ and\ \citenamefont {Firsov}}]{Novoselov2004}%
  \BibitemOpen
  \bibfield  {author} {\bibinfo {author} {\bibfnamefont {K.~S.}\ \bibnamefont
  {Novoselov}}, \bibinfo {author} {\bibfnamefont {A.~K.}\ \bibnamefont {Geim}},
  \bibinfo {author} {\bibfnamefont {S.~V.}\ \bibnamefont {Morozov}}, \bibinfo
  {author} {\bibfnamefont {D.}~\bibnamefont {Jiang}}, \bibinfo {author}
  {\bibfnamefont {Y.}~\bibnamefont {Zhang}}, \bibinfo {author} {\bibfnamefont
  {S.~V.}\ \bibnamefont {Dubonos}}, \bibinfo {author} {\bibfnamefont {I.~V.}\
  \bibnamefont {Grigorieva}}, \ and\ \bibinfo {author} {\bibfnamefont {A.~A.}\
  \bibnamefont {Firsov}},\ }\href {\doibase 10.1126/science.1102896} {\bibfield
   {journal} {\bibinfo  {journal} {Science}\ }\textbf {\bibinfo {volume}
  {306}},\ \bibinfo {pages} {666} (\bibinfo {year} {2004})}\BibitemShut
  {NoStop}%
\bibitem [{\citenamefont {Song}\ \emph {et~al.}(2010)\citenamefont {Song},
  \citenamefont {Ci}, \citenamefont {Lu}, \citenamefont {Sorokin},
  \citenamefont {Jin}, \citenamefont {Ni}, \citenamefont {Kvashnin},
  \citenamefont {Kvashnin}, \citenamefont {Lou}, \citenamefont {Yakobson},\
  and\ \citenamefont {Ajayan}}]{Song2010}%
  \BibitemOpen
  \bibfield  {author} {\bibinfo {author} {\bibfnamefont {L.}~\bibnamefont
  {Song}}, \bibinfo {author} {\bibfnamefont {L.}~\bibnamefont {Ci}}, \bibinfo
  {author} {\bibfnamefont {H.}~\bibnamefont {Lu}}, \bibinfo {author}
  {\bibfnamefont {P.~B.}\ \bibnamefont {Sorokin}}, \bibinfo {author}
  {\bibfnamefont {C.}~\bibnamefont {Jin}}, \bibinfo {author} {\bibfnamefont
  {J.}~\bibnamefont {Ni}}, \bibinfo {author} {\bibfnamefont {A.~G.}\
  \bibnamefont {Kvashnin}}, \bibinfo {author} {\bibfnamefont {D.~G.}\
  \bibnamefont {Kvashnin}}, \bibinfo {author} {\bibfnamefont {J.}~\bibnamefont
  {Lou}}, \bibinfo {author} {\bibfnamefont {B.~I.}\ \bibnamefont {Yakobson}}, \
  and\ \bibinfo {author} {\bibfnamefont {P.~M.}\ \bibnamefont {Ajayan}},\
  }\href {\doibase 10.1021/nl1022139} {\bibfield  {journal} {\bibinfo
  {journal} {Nano Letters}\ }\textbf {\bibinfo {volume} {10}},\ \bibinfo
  {pages} {3209} (\bibinfo {year} {2010})}\BibitemShut {NoStop}%
\bibitem [{\citenamefont {Lalmi}\ \emph {et~al.}(2010)\citenamefont {Lalmi},
  \citenamefont {Oughaddou}, \citenamefont {Enriquez}, \citenamefont {Kara},
  \citenamefont {Vizzini}, \citenamefont {Ealet},\ and\ \citenamefont
  {Aufray}}]{Lalmi2010}%
  \BibitemOpen
  \bibfield  {author} {\bibinfo {author} {\bibfnamefont {B.}~\bibnamefont
  {Lalmi}}, \bibinfo {author} {\bibfnamefont {H.}~\bibnamefont {Oughaddou}},
  \bibinfo {author} {\bibfnamefont {H.}~\bibnamefont {Enriquez}}, \bibinfo
  {author} {\bibfnamefont {A.}~\bibnamefont {Kara}}, \bibinfo {author}
  {\bibfnamefont {S.}~\bibnamefont {Vizzini}}, \bibinfo {author} {\bibfnamefont
  {B.}~\bibnamefont {Ealet}}, \ and\ \bibinfo {author} {\bibfnamefont
  {B.}~\bibnamefont {Aufray}},\ }\href {\doibase 10.1063/1.3524215} {\bibfield
  {journal} {\bibinfo  {journal} {Applied Physics Letters}\ }\textbf {\bibinfo
  {volume} {97}},\ \bibinfo {pages} {223109} (\bibinfo {year}
  {2010})}\BibitemShut {NoStop}%
\bibitem [{\citenamefont {Mannix}\ \emph {et~al.}(2015)\citenamefont {Mannix},
  \citenamefont {Zhou}, \citenamefont {Kiraly}, \citenamefont {Wood},
  \citenamefont {Alducin}, \citenamefont {Myers}, \citenamefont {Liu},
  \citenamefont {Fisher}, \citenamefont {Santiago}, \citenamefont {Guest},
  \citenamefont {Yacaman}, \citenamefont {Ponce}, \citenamefont {Oganov},
  \citenamefont {Hersam},\ and\ \citenamefont {Guisinger}}]{Mannix2015}%
  \BibitemOpen
  \bibfield  {author} {\bibinfo {author} {\bibfnamefont {A.~J.}\ \bibnamefont
  {Mannix}}, \bibinfo {author} {\bibfnamefont {X.-F.}\ \bibnamefont {Zhou}},
  \bibinfo {author} {\bibfnamefont {B.}~\bibnamefont {Kiraly}}, \bibinfo
  {author} {\bibfnamefont {J.~D.}\ \bibnamefont {Wood}}, \bibinfo {author}
  {\bibfnamefont {D.}~\bibnamefont {Alducin}}, \bibinfo {author} {\bibfnamefont
  {B.~D.}\ \bibnamefont {Myers}}, \bibinfo {author} {\bibfnamefont
  {X.}~\bibnamefont {Liu}}, \bibinfo {author} {\bibfnamefont {B.~L.}\
  \bibnamefont {Fisher}}, \bibinfo {author} {\bibfnamefont {U.}~\bibnamefont
  {Santiago}}, \bibinfo {author} {\bibfnamefont {J.~R.}\ \bibnamefont {Guest}},
  \bibinfo {author} {\bibfnamefont {M.~J.}\ \bibnamefont {Yacaman}}, \bibinfo
  {author} {\bibfnamefont {A.}~\bibnamefont {Ponce}}, \bibinfo {author}
  {\bibfnamefont {A.~R.}\ \bibnamefont {Oganov}}, \bibinfo {author}
  {\bibfnamefont {M.~C.}\ \bibnamefont {Hersam}}, \ and\ \bibinfo {author}
  {\bibfnamefont {N.~P.}\ \bibnamefont {Guisinger}},\ }\href {\doibase
  10.1126/science.aad1080} {\bibfield  {journal} {\bibinfo  {journal}
  {Science}\ }\textbf {\bibinfo {volume} {350}},\ \bibinfo {pages} {1513}
  (\bibinfo {year} {2015})}\BibitemShut {NoStop}%
\bibitem [{\citenamefont {Saxena}\ \emph {et~al.}(2016)\citenamefont {Saxena},
  \citenamefont {Chaudhary},\ and\ \citenamefont {Shukla}}]{Saxena2016}%
  \BibitemOpen
  \bibfield  {author} {\bibinfo {author} {\bibfnamefont {S.}~\bibnamefont
  {Saxena}}, \bibinfo {author} {\bibfnamefont {R.~P.}\ \bibnamefont
  {Chaudhary}}, \ and\ \bibinfo {author} {\bibfnamefont {S.}~\bibnamefont
  {Shukla}},\ }\href {\doibase 10.1038/srep31073} {\bibfield  {journal}
  {\bibinfo  {journal} {Scientific Reports}\ }\textbf {\bibinfo {volume} {6}},\
  \bibinfo {pages} {31073} (\bibinfo {year} {2016})}\BibitemShut {NoStop}%
\bibitem [{\citenamefont {Coleman}\ \emph {et~al.}(2011)\citenamefont
  {Coleman}, \citenamefont {Lotya}, \citenamefont {O'Neill}, \citenamefont
  {Bergin}, \citenamefont {King}, \citenamefont {Khan}, \citenamefont {Young},
  \citenamefont {Gaucher}, \citenamefont {De}, \citenamefont {Smith},
  \citenamefont {Shvets}, \citenamefont {Arora}, \citenamefont {Stanton},
  \citenamefont {Kim}, \citenamefont {Lee}, \citenamefont {Kim}, \citenamefont
  {Duesberg}, \citenamefont {Hallam}, \citenamefont {Boland}, \citenamefont
  {Wang}, \citenamefont {Donegan}, \citenamefont {Grunlan}, \citenamefont
  {Moriarty}, \citenamefont {Shmeliov}, \citenamefont {Nicholls}, \citenamefont
  {Perkins}, \citenamefont {Grieveson}, \citenamefont {Theuwissen},
  \citenamefont {McComb}, \citenamefont {Nellist},\ and\ \citenamefont
  {Nicolosi}}]{Coleman2011}%
  \BibitemOpen
  \bibfield  {author} {\bibinfo {author} {\bibfnamefont {J.~N.}\ \bibnamefont
  {Coleman}}, \bibinfo {author} {\bibfnamefont {M.}~\bibnamefont {Lotya}},
  \bibinfo {author} {\bibfnamefont {A.}~\bibnamefont {O'Neill}}, \bibinfo
  {author} {\bibfnamefont {S.~D.}\ \bibnamefont {Bergin}}, \bibinfo {author}
  {\bibfnamefont {P.~J.}\ \bibnamefont {King}}, \bibinfo {author}
  {\bibfnamefont {U.}~\bibnamefont {Khan}}, \bibinfo {author} {\bibfnamefont
  {K.}~\bibnamefont {Young}}, \bibinfo {author} {\bibfnamefont
  {A.}~\bibnamefont {Gaucher}}, \bibinfo {author} {\bibfnamefont
  {S.}~\bibnamefont {De}}, \bibinfo {author} {\bibfnamefont {R.~J.}\
  \bibnamefont {Smith}}, \bibinfo {author} {\bibfnamefont {I.~V.}\ \bibnamefont
  {Shvets}}, \bibinfo {author} {\bibfnamefont {S.~K.}\ \bibnamefont {Arora}},
  \bibinfo {author} {\bibfnamefont {G.}~\bibnamefont {Stanton}}, \bibinfo
  {author} {\bibfnamefont {H.-Y.}\ \bibnamefont {Kim}}, \bibinfo {author}
  {\bibfnamefont {K.}~\bibnamefont {Lee}}, \bibinfo {author} {\bibfnamefont
  {G.~T.}\ \bibnamefont {Kim}}, \bibinfo {author} {\bibfnamefont {G.~S.}\
  \bibnamefont {Duesberg}}, \bibinfo {author} {\bibfnamefont {T.}~\bibnamefont
  {Hallam}}, \bibinfo {author} {\bibfnamefont {J.~J.}\ \bibnamefont {Boland}},
  \bibinfo {author} {\bibfnamefont {J.~J.}\ \bibnamefont {Wang}}, \bibinfo
  {author} {\bibfnamefont {J.~F.}\ \bibnamefont {Donegan}}, \bibinfo {author}
  {\bibfnamefont {J.~C.}\ \bibnamefont {Grunlan}}, \bibinfo {author}
  {\bibfnamefont {G.}~\bibnamefont {Moriarty}}, \bibinfo {author}
  {\bibfnamefont {A.}~\bibnamefont {Shmeliov}}, \bibinfo {author}
  {\bibfnamefont {R.~J.}\ \bibnamefont {Nicholls}}, \bibinfo {author}
  {\bibfnamefont {J.~M.}\ \bibnamefont {Perkins}}, \bibinfo {author}
  {\bibfnamefont {E.~M.}\ \bibnamefont {Grieveson}}, \bibinfo {author}
  {\bibfnamefont {K.}~\bibnamefont {Theuwissen}}, \bibinfo {author}
  {\bibfnamefont {D.~W.}\ \bibnamefont {McComb}}, \bibinfo {author}
  {\bibfnamefont {P.~D.}\ \bibnamefont {Nellist}}, \ and\ \bibinfo {author}
  {\bibfnamefont {V.}~\bibnamefont {Nicolosi}},\ }\href {\doibase
  10.1126/science.1194975} {\bibfield  {journal} {\bibinfo  {journal}
  {Science}\ }\textbf {\bibinfo {volume} {331}},\ \bibinfo {pages} {568}
  (\bibinfo {year} {2011})}\BibitemShut {NoStop}%
\bibitem [{\citenamefont {Naguib}\ \emph {et~al.}(2011)\citenamefont {Naguib},
  \citenamefont {Kurtoglu}, \citenamefont {Presser}, \citenamefont {Lu},
  \citenamefont {Niu}, \citenamefont {Heon}, \citenamefont {Hultman},
  \citenamefont {Gogotsi},\ and\ \citenamefont {Barsoum}}]{Naguib2011}%
  \BibitemOpen
  \bibfield  {author} {\bibinfo {author} {\bibfnamefont {M.}~\bibnamefont
  {Naguib}}, \bibinfo {author} {\bibfnamefont {M.}~\bibnamefont {Kurtoglu}},
  \bibinfo {author} {\bibfnamefont {V.}~\bibnamefont {Presser}}, \bibinfo
  {author} {\bibfnamefont {J.}~\bibnamefont {Lu}}, \bibinfo {author}
  {\bibfnamefont {J.}~\bibnamefont {Niu}}, \bibinfo {author} {\bibfnamefont
  {M.}~\bibnamefont {Heon}}, \bibinfo {author} {\bibfnamefont {L.}~\bibnamefont
  {Hultman}}, \bibinfo {author} {\bibfnamefont {Y.}~\bibnamefont {Gogotsi}}, \
  and\ \bibinfo {author} {\bibfnamefont {M.~W.}\ \bibnamefont {Barsoum}},\
  }\href {\doibase 10.1002/adma.201102306} {\bibfield  {journal} {\bibinfo
  {journal} {Advanced Materials}\ }\textbf {\bibinfo {volume} {23}},\ \bibinfo
  {pages} {4248} (\bibinfo {year} {2011})}\BibitemShut {NoStop}%
\bibitem [{\citenamefont {Ashton}\ \emph {et~al.}(2017)\citenamefont {Ashton},
  \citenamefont {Paul}, \citenamefont {Sinnott},\ and\ \citenamefont
  {Hennig}}]{Ashton2017}%
  \BibitemOpen
  \bibfield  {author} {\bibinfo {author} {\bibfnamefont {M.}~\bibnamefont
  {Ashton}}, \bibinfo {author} {\bibfnamefont {J.}~\bibnamefont {Paul}},
  \bibinfo {author} {\bibfnamefont {S.~B.}\ \bibnamefont {Sinnott}}, \ and\
  \bibinfo {author} {\bibfnamefont {R.~G.}\ \bibnamefont {Hennig}},\ }\href
  {\doibase 10.1103/PhysRevLett.118.106101} {\bibfield  {journal} {\bibinfo
  {journal} {Physical Review Letters}\ }\textbf {\bibinfo {volume} {118}},\
  \bibinfo {pages} {106101} (\bibinfo {year} {2017})}\BibitemShut {NoStop}%
\bibitem [{\citenamefont {Hu}\ \emph {et~al.}(2021)\citenamefont {Hu},
  \citenamefont {Wang}, \citenamefont {Tan}, \citenamefont {Duan},
  \citenamefont {Li}, \citenamefont {Li}, \citenamefont {Ji}, \citenamefont
  {Lu}, \citenamefont {Wang}, \citenamefont {Sun}, \citenamefont {Hu},\ and\
  \citenamefont {Yan}}]{Hu2021}%
  \BibitemOpen
  \bibfield  {author} {\bibinfo {author} {\bibfnamefont {W.}~\bibnamefont
  {Hu}}, \bibinfo {author} {\bibfnamefont {C.}~\bibnamefont {Wang}}, \bibinfo
  {author} {\bibfnamefont {H.}~\bibnamefont {Tan}}, \bibinfo {author}
  {\bibfnamefont {H.}~\bibnamefont {Duan}}, \bibinfo {author} {\bibfnamefont
  {G.}~\bibnamefont {Li}}, \bibinfo {author} {\bibfnamefont {N.}~\bibnamefont
  {Li}}, \bibinfo {author} {\bibfnamefont {Q.}~\bibnamefont {Ji}}, \bibinfo
  {author} {\bibfnamefont {Y.}~\bibnamefont {Lu}}, \bibinfo {author}
  {\bibfnamefont {Y.}~\bibnamefont {Wang}}, \bibinfo {author} {\bibfnamefont
  {Z.}~\bibnamefont {Sun}}, \bibinfo {author} {\bibfnamefont {F.}~\bibnamefont
  {Hu}}, \ and\ \bibinfo {author} {\bibfnamefont {W.}~\bibnamefont {Yan}},\
  }\href {\doibase 10.1038/s41467-021-22122-2} {\bibfield  {journal} {\bibinfo
  {journal} {Nature Communications}\ }\textbf {\bibinfo {volume} {12}},\
  \bibinfo {pages} {1854} (\bibinfo {year} {2021})}\BibitemShut {NoStop}%
\bibitem [{\citenamefont {He}\ \emph {et~al.}(2019)\citenamefont {He},
  \citenamefont {He}, \citenamefont {Deng}, \citenamefont {Peng}, \citenamefont
  {Chen}, \citenamefont {Zhang}, \citenamefont {Yao}, \citenamefont {Zhang},
  \citenamefont {Xiao}, \citenamefont {Ma}, \citenamefont {Ge},\ and\
  \citenamefont {Ji}}]{He2019}%
  \BibitemOpen
  \bibfield  {author} {\bibinfo {author} {\bibfnamefont {X.}~\bibnamefont
  {He}}, \bibinfo {author} {\bibfnamefont {Q.}~\bibnamefont {He}}, \bibinfo
  {author} {\bibfnamefont {Y.}~\bibnamefont {Deng}}, \bibinfo {author}
  {\bibfnamefont {M.}~\bibnamefont {Peng}}, \bibinfo {author} {\bibfnamefont
  {H.}~\bibnamefont {Chen}}, \bibinfo {author} {\bibfnamefont {Y.}~\bibnamefont
  {Zhang}}, \bibinfo {author} {\bibfnamefont {S.}~\bibnamefont {Yao}}, \bibinfo
  {author} {\bibfnamefont {M.}~\bibnamefont {Zhang}}, \bibinfo {author}
  {\bibfnamefont {D.}~\bibnamefont {Xiao}}, \bibinfo {author} {\bibfnamefont
  {D.}~\bibnamefont {Ma}}, \bibinfo {author} {\bibfnamefont {B.}~\bibnamefont
  {Ge}}, \ and\ \bibinfo {author} {\bibfnamefont {H.}~\bibnamefont {Ji}},\
  }\href {\doibase 10.1038/s41467-019-11619-6} {\bibfield  {journal} {\bibinfo
  {journal} {Nature Communications}\ }\textbf {\bibinfo {volume} {10}},\
  \bibinfo {pages} {3663} (\bibinfo {year} {2019})}\BibitemShut {NoStop}%
\bibitem [{\citenamefont {Lai}\ \emph {et~al.}(2020)\citenamefont {Lai},
  \citenamefont {Wang}, \citenamefont {Zheng}, \citenamefont {Jiang},
  \citenamefont {Yan}, \citenamefont {Wang}, \citenamefont {Yoshikawa},
  \citenamefont {Matsumura}, \citenamefont {Sun}, \citenamefont {Wang},
  \citenamefont {Gu}, \citenamefont {Wang}, \citenamefont {Liu}, \citenamefont
  {Chou},\ and\ \citenamefont {Dou}}]{LAI2020}%
  \BibitemOpen
  \bibfield  {author} {\bibinfo {author} {\bibfnamefont {W.~H.}\ \bibnamefont
  {Lai}}, \bibinfo {author} {\bibfnamefont {H.}~\bibnamefont {Wang}}, \bibinfo
  {author} {\bibfnamefont {L.}~\bibnamefont {Zheng}}, \bibinfo {author}
  {\bibfnamefont {Q.}~\bibnamefont {Jiang}}, \bibinfo {author} {\bibfnamefont
  {Z.~C.}\ \bibnamefont {Yan}}, \bibinfo {author} {\bibfnamefont
  {L.}~\bibnamefont {Wang}}, \bibinfo {author} {\bibfnamefont {H.}~\bibnamefont
  {Yoshikawa}}, \bibinfo {author} {\bibfnamefont {D.}~\bibnamefont
  {Matsumura}}, \bibinfo {author} {\bibfnamefont {Q.}~\bibnamefont {Sun}},
  \bibinfo {author} {\bibfnamefont {Y.~X.}\ \bibnamefont {Wang}}, \bibinfo
  {author} {\bibfnamefont {Q.}~\bibnamefont {Gu}}, \bibinfo {author}
  {\bibfnamefont {J.~Z.}\ \bibnamefont {Wang}}, \bibinfo {author}
  {\bibfnamefont {H.~K.}\ \bibnamefont {Liu}}, \bibinfo {author} {\bibfnamefont
  {S.~L.}\ \bibnamefont {Chou}}, \ and\ \bibinfo {author} {\bibfnamefont
  {S.~X.}\ \bibnamefont {Dou}},\ }\href {\doibase 10.1002/anie.202009400}
  {\bibfield  {journal} {\bibinfo  {journal} {Angewandte Chemie - International
  Edition}\ }\textbf {\bibinfo {volume} {59}},\ \bibinfo {pages} {22171}
  (\bibinfo {year} {2020})}\BibitemShut {NoStop}%
\bibitem [{\citenamefont {Fei}\ \emph {et~al.}(2018)\citenamefont {Fei},
  \citenamefont {Dong}, \citenamefont {Feng}, \citenamefont {Allen},
  \citenamefont {Wan}, \citenamefont {Volosskiy}, \citenamefont {Li},
  \citenamefont {Zhao}, \citenamefont {Wang}, \citenamefont {Sun},
  \citenamefont {An}, \citenamefont {Chen}, \citenamefont {Guo}, \citenamefont
  {Lee}, \citenamefont {Chen}, \citenamefont {Shakir}, \citenamefont {Liu},
  \citenamefont {Hu}, \citenamefont {Li}, \citenamefont {Kirkland},
  \citenamefont {Duan},\ and\ \citenamefont {Huang}}]{Fei2018}%
  \BibitemOpen
  \bibfield  {author} {\bibinfo {author} {\bibfnamefont {H.}~\bibnamefont
  {Fei}}, \bibinfo {author} {\bibfnamefont {J.}~\bibnamefont {Dong}}, \bibinfo
  {author} {\bibfnamefont {Y.}~\bibnamefont {Feng}}, \bibinfo {author}
  {\bibfnamefont {C.~S.}\ \bibnamefont {Allen}}, \bibinfo {author}
  {\bibfnamefont {C.}~\bibnamefont {Wan}}, \bibinfo {author} {\bibfnamefont
  {B.}~\bibnamefont {Volosskiy}}, \bibinfo {author} {\bibfnamefont
  {M.}~\bibnamefont {Li}}, \bibinfo {author} {\bibfnamefont {Z.}~\bibnamefont
  {Zhao}}, \bibinfo {author} {\bibfnamefont {Y.}~\bibnamefont {Wang}}, \bibinfo
  {author} {\bibfnamefont {H.}~\bibnamefont {Sun}}, \bibinfo {author}
  {\bibfnamefont {P.}~\bibnamefont {An}}, \bibinfo {author} {\bibfnamefont
  {W.}~\bibnamefont {Chen}}, \bibinfo {author} {\bibfnamefont {Z.}~\bibnamefont
  {Guo}}, \bibinfo {author} {\bibfnamefont {C.}~\bibnamefont {Lee}}, \bibinfo
  {author} {\bibfnamefont {D.}~\bibnamefont {Chen}}, \bibinfo {author}
  {\bibfnamefont {I.}~\bibnamefont {Shakir}}, \bibinfo {author} {\bibfnamefont
  {M.}~\bibnamefont {Liu}}, \bibinfo {author} {\bibfnamefont {T.}~\bibnamefont
  {Hu}}, \bibinfo {author} {\bibfnamefont {Y.}~\bibnamefont {Li}}, \bibinfo
  {author} {\bibfnamefont {A.~I.}\ \bibnamefont {Kirkland}}, \bibinfo {author}
  {\bibfnamefont {X.}~\bibnamefont {Duan}}, \ and\ \bibinfo {author}
  {\bibfnamefont {Y.}~\bibnamefont {Huang}},\ }\href {\doibase
  10.1038/s41929-017-0008-y} {\bibfield  {journal} {\bibinfo  {journal} {Nature
  Catalysis}\ }\textbf {\bibinfo {volume} {1}},\ \bibinfo {pages} {63}
  (\bibinfo {year} {2018})}\BibitemShut {NoStop}%
\bibitem [{\citenamefont {Liu}\ \emph {et~al.}(2021{\natexlab{a}})\citenamefont
  {Liu}, \citenamefont {Zhang}, \citenamefont {Gao},\ and\ \citenamefont
  {Yan}}]{Liu2021}%
  \BibitemOpen
  \bibfield  {author} {\bibinfo {author} {\bibfnamefont {D.}~\bibnamefont
  {Liu}}, \bibinfo {author} {\bibfnamefont {S.}~\bibnamefont {Zhang}}, \bibinfo
  {author} {\bibfnamefont {M.}~\bibnamefont {Gao}}, \ and\ \bibinfo {author}
  {\bibfnamefont {X.-W.}\ \bibnamefont {Yan}},\ }\href {\doibase
  10.1103/PhysRevB.103.125407} {\bibfield  {journal} {\bibinfo  {journal}
  {Physical Review B}\ }\textbf {\bibinfo {volume} {103}},\ \bibinfo {pages}
  {125407} (\bibinfo {year} {2021}{\natexlab{a}})}\BibitemShut {NoStop}%
\bibitem [{\citenamefont {Liu}\ \emph {et~al.}(2021{\natexlab{b}})\citenamefont
  {Liu}, \citenamefont {Feng}, \citenamefont {Gao},\ and\ \citenamefont
  {Yan}}]{Liu2021a}%
  \BibitemOpen
  \bibfield  {author} {\bibinfo {author} {\bibfnamefont {D.}~\bibnamefont
  {Liu}}, \bibinfo {author} {\bibfnamefont {P.}~\bibnamefont {Feng}}, \bibinfo
  {author} {\bibfnamefont {M.}~\bibnamefont {Gao}}, \ and\ \bibinfo {author}
  {\bibfnamefont {X.-W.}\ \bibnamefont {Yan}},\ }\href {\doibase
  10.1103/PhysRevB.103.155411} {\bibfield  {journal} {\bibinfo  {journal}
  {Physical Review B}\ }\textbf {\bibinfo {volume} {103}},\ \bibinfo {pages}
  {155411} (\bibinfo {year} {2021}{\natexlab{b}})}\BibitemShut {NoStop}%
\bibitem [{\citenamefont {Liu}\ \emph {et~al.}(2021{\natexlab{c}})\citenamefont
  {Liu}, \citenamefont {Zhang}, \citenamefont {Gao}, \citenamefont {Yan},\ and\
  \citenamefont {Xie}}]{Liu2021b}%
  \BibitemOpen
  \bibfield  {author} {\bibinfo {author} {\bibfnamefont {D.}~\bibnamefont
  {Liu}}, \bibinfo {author} {\bibfnamefont {S.}~\bibnamefont {Zhang}}, \bibinfo
  {author} {\bibfnamefont {M.}~\bibnamefont {Gao}}, \bibinfo {author}
  {\bibfnamefont {X.-W.}\ \bibnamefont {Yan}}, \ and\ \bibinfo {author}
  {\bibfnamefont {Z.~Y.}\ \bibnamefont {Xie}},\ }\href {\doibase
  10.1063/5.0054730} {\bibfield  {journal} {\bibinfo  {journal} {Applied
  Physics Letters}\ }\textbf {\bibinfo {volume} {118}},\ \bibinfo {pages}
  {223104} (\bibinfo {year} {2021}{\natexlab{c}})}\BibitemShut {NoStop}%
\bibitem [{\citenamefont {Bykov}\ \emph {et~al.}(2021)\citenamefont {Bykov},
  \citenamefont {Fedotenko}, \citenamefont {Chariton}, \citenamefont {Laniel},
  \citenamefont {Glazyrin}, \citenamefont {Hanfland}, \citenamefont {Smith},
  \citenamefont {Prakapenka}, \citenamefont {Mahmood}, \citenamefont
  {Goncharov}, \citenamefont {Ponomareva}, \citenamefont {Tasn{\'{a}}di},
  \citenamefont {Abrikosov}, \citenamefont {{Bin Masood}}, \citenamefont
  {Hotz}, \citenamefont {Rudenko}, \citenamefont {Katsnelson}, \citenamefont
  {Dubrovinskaia}, \citenamefont {Dubrovinsky},\ and\ \citenamefont
  {Abrikosov}}]{Bykov2021}%
  \BibitemOpen
  \bibfield  {author} {\bibinfo {author} {\bibfnamefont {M.}~\bibnamefont
  {Bykov}}, \bibinfo {author} {\bibfnamefont {T.}~\bibnamefont {Fedotenko}},
  \bibinfo {author} {\bibfnamefont {S.}~\bibnamefont {Chariton}}, \bibinfo
  {author} {\bibfnamefont {D.}~\bibnamefont {Laniel}}, \bibinfo {author}
  {\bibfnamefont {K.}~\bibnamefont {Glazyrin}}, \bibinfo {author}
  {\bibfnamefont {M.}~\bibnamefont {Hanfland}}, \bibinfo {author}
  {\bibfnamefont {J.~S.}\ \bibnamefont {Smith}}, \bibinfo {author}
  {\bibfnamefont {V.~B.}\ \bibnamefont {Prakapenka}}, \bibinfo {author}
  {\bibfnamefont {M.~F.}\ \bibnamefont {Mahmood}}, \bibinfo {author}
  {\bibfnamefont {A.~F.}\ \bibnamefont {Goncharov}}, \bibinfo {author}
  {\bibfnamefont {A.~V.}\ \bibnamefont {Ponomareva}}, \bibinfo {author}
  {\bibfnamefont {F.}~\bibnamefont {Tasn{\'{a}}di}}, \bibinfo {author}
  {\bibfnamefont {A.~I.}\ \bibnamefont {Abrikosov}}, \bibinfo {author}
  {\bibfnamefont {T.}~\bibnamefont {{Bin Masood}}}, \bibinfo {author}
  {\bibfnamefont {I.}~\bibnamefont {Hotz}}, \bibinfo {author} {\bibfnamefont
  {A.~N.}\ \bibnamefont {Rudenko}}, \bibinfo {author} {\bibfnamefont {M.~I.}\
  \bibnamefont {Katsnelson}}, \bibinfo {author} {\bibfnamefont
  {N.}~\bibnamefont {Dubrovinskaia}}, \bibinfo {author} {\bibfnamefont
  {L.}~\bibnamefont {Dubrovinsky}}, \ and\ \bibinfo {author} {\bibfnamefont
  {I.~A.}\ \bibnamefont {Abrikosov}},\ }\href {\doibase
  10.1103/PhysRevLett.126.175501} {\bibfield  {journal} {\bibinfo  {journal}
  {Physical Review Letters}\ }\textbf {\bibinfo {volume} {126}},\ \bibinfo
  {pages} {175501} (\bibinfo {year} {2021})}\BibitemShut {NoStop}%
\bibitem [{\citenamefont {Mortazavi}\ \emph {et~al.}(2021)\citenamefont
  {Mortazavi}, \citenamefont {Shojaei},\ and\ \citenamefont
  {Zhuang}}]{Mortazavi2021}%
  \BibitemOpen
  \bibfield  {author} {\bibinfo {author} {\bibfnamefont {B.}~\bibnamefont
  {Mortazavi}}, \bibinfo {author} {\bibfnamefont {F.}~\bibnamefont {Shojaei}},
  \ and\ \bibinfo {author} {\bibfnamefont {X.}~\bibnamefont {Zhuang}},\ }\href
  {\doibase 10.1016/j.mtnano.2021.100125} {\bibfield  {journal} {\bibinfo
  {journal} {Materials Today Nano}\ }\textbf {\bibinfo {volume} {15}},\
  \bibinfo {pages} {2} (\bibinfo {year} {2021})}\BibitemShut {NoStop}%
\bibitem [{\citenamefont {Zhang}\ \emph {et~al.}(2015)\citenamefont {Zhang},
  \citenamefont {Li}, \citenamefont {Zhao},\ and\ \citenamefont
  {Wang}}]{Zhang2015}%
  \BibitemOpen
  \bibfield  {author} {\bibinfo {author} {\bibfnamefont {S.}~\bibnamefont
  {Zhang}}, \bibinfo {author} {\bibfnamefont {Y.}~\bibnamefont {Li}}, \bibinfo
  {author} {\bibfnamefont {T.}~\bibnamefont {Zhao}}, \ and\ \bibinfo {author}
  {\bibfnamefont {Q.}~\bibnamefont {Wang}},\ }\href {\doibase
  10.1038/srep05241} {\bibfield  {journal} {\bibinfo  {journal} {Scientific
  Reports}\ }\textbf {\bibinfo {volume} {4}},\ \bibinfo {pages} {5241}
  (\bibinfo {year} {2015})}\BibitemShut {NoStop}%
\bibitem [{\citenamefont {Tang}\ \emph {et~al.}(2019)\citenamefont {Tang},
  \citenamefont {Kour},\ and\ \citenamefont {Du}}]{Tang2019}%
  \BibitemOpen
  \bibfield  {author} {\bibinfo {author} {\bibfnamefont {C.}~\bibnamefont
  {Tang}}, \bibinfo {author} {\bibfnamefont {G.}~\bibnamefont {Kour}}, \ and\
  \bibinfo {author} {\bibfnamefont {A.}~\bibnamefont {Du}},\ }\href {\doibase
  10.1088/1674-1056/ab41ea} {\bibfield  {journal} {\bibinfo  {journal} {Chinese
  Physics B}\ }\textbf {\bibinfo {volume} {28}},\ \bibinfo {pages} {107306}
  (\bibinfo {year} {2019})}\BibitemShut {NoStop}%
\bibitem [{\citenamefont {Kresse}\ and\ \citenamefont
  {Hafner}(1993)}]{PhysRevB.47.558}%
  \BibitemOpen
  \bibfield  {author} {\bibinfo {author} {\bibfnamefont {G.}~\bibnamefont
  {Kresse}}\ and\ \bibinfo {author} {\bibfnamefont {J.}~\bibnamefont
  {Hafner}},\ }\href {\doibase 10.1103/PhysRevB.47.558} {\bibfield  {journal}
  {\bibinfo  {journal} {Phys. Rev. B}\ }\textbf {\bibinfo {volume} {47}},\
  \bibinfo {pages} {558} (\bibinfo {year} {1993})}\BibitemShut {NoStop}%
\bibitem [{\citenamefont {Kresse}\ and\ \citenamefont
  {Furthm{\"{u}}ller}(1996)}]{PhysRevB.54.11169}%
  \BibitemOpen
  \bibfield  {author} {\bibinfo {author} {\bibfnamefont {G.}~\bibnamefont
  {Kresse}}\ and\ \bibinfo {author} {\bibfnamefont {J.}~\bibnamefont
  {Furthm{\"{u}}ller}},\ }\href {\doibase 10.1103/PhysRevB.54.11169} {\bibfield
   {journal} {\bibinfo  {journal} {Physical Review B}\ }\textbf {\bibinfo
  {volume} {54}},\ \bibinfo {pages} {11169} (\bibinfo {year}
  {1996})}\BibitemShut {NoStop}%
\bibitem [{\citenamefont {Perdew}\ \emph {et~al.}(1996)\citenamefont {Perdew},
  \citenamefont {Burke},\ and\ \citenamefont
  {Ernzerhof}}]{PhysRevLett.77.3865}%
  \BibitemOpen
  \bibfield  {author} {\bibinfo {author} {\bibfnamefont {J.~P.}\ \bibnamefont
  {Perdew}}, \bibinfo {author} {\bibfnamefont {K.}~\bibnamefont {Burke}}, \
  and\ \bibinfo {author} {\bibfnamefont {M.}~\bibnamefont {Ernzerhof}},\ }\href
  {\doibase 10.1103/PhysRevLett.77.3865} {\bibfield  {journal} {\bibinfo
  {journal} {Physical Review Letters}\ }\textbf {\bibinfo {volume} {77}},\
  \bibinfo {pages} {3865} (\bibinfo {year} {1996})}\BibitemShut {NoStop}%
\bibitem [{\citenamefont {Bl{\"{o}}chl}(1994)}]{PhysRevB.50.17953}%
  \BibitemOpen
  \bibfield  {author} {\bibinfo {author} {\bibfnamefont {P.~E.}\ \bibnamefont
  {Bl{\"{o}}chl}},\ }\href {\doibase 10.1103/PhysRevB.50.17953} {\bibfield
  {journal} {\bibinfo  {journal} {Physical Review B}\ }\textbf {\bibinfo
  {volume} {50}},\ \bibinfo {pages} {17953} (\bibinfo {year}
  {1994})}\BibitemShut {NoStop}%
\bibitem [{\citenamefont {Sun}\ \emph {et~al.}(2015)\citenamefont {Sun},
  \citenamefont {Ruzsinszky},\ and\ \citenamefont {Perdew}}]{Sun2015}%
  \BibitemOpen
  \bibfield  {author} {\bibinfo {author} {\bibfnamefont {J.}~\bibnamefont
  {Sun}}, \bibinfo {author} {\bibfnamefont {A.}~\bibnamefont {Ruzsinszky}}, \
  and\ \bibinfo {author} {\bibfnamefont {J.~P.}\ \bibnamefont {Perdew}},\
  }\href {\doibase 10.1103/PhysRevLett.115.036402} {\bibfield  {journal}
  {\bibinfo  {journal} {Physical Review Letters}\ }\textbf {\bibinfo {volume}
  {115}},\ \bibinfo {pages} {036402} (\bibinfo {year} {2015})}\BibitemShut
  {NoStop}%
\bibitem [{\citenamefont {Cococcioni}\ and\ \citenamefont
  {de~Gironcoli}(2005)}]{Cococcioni2005}%
  \BibitemOpen
  \bibfield  {author} {\bibinfo {author} {\bibfnamefont {M.}~\bibnamefont
  {Cococcioni}}\ and\ \bibinfo {author} {\bibfnamefont {S.}~\bibnamefont
  {de~Gironcoli}},\ }\href {\doibase 10.1103/PhysRevB.71.035105} {\bibfield
  {journal} {\bibinfo  {journal} {Physical Review B}\ }\textbf {\bibinfo
  {volume} {71}},\ \bibinfo {pages} {035105} (\bibinfo {year}
  {2005})}\BibitemShut {NoStop}%
\bibitem [{\citenamefont {Krukau}\ \emph {et~al.}(2006)\citenamefont {Krukau},
  \citenamefont {Vydrov}, \citenamefont {Izmaylov},\ and\ \citenamefont
  {Scuseria}}]{Krukau2006}%
  \BibitemOpen
  \bibfield  {author} {\bibinfo {author} {\bibfnamefont {A.~V.}\ \bibnamefont
  {Krukau}}, \bibinfo {author} {\bibfnamefont {O.~A.}\ \bibnamefont {Vydrov}},
  \bibinfo {author} {\bibfnamefont {A.~F.}\ \bibnamefont {Izmaylov}}, \ and\
  \bibinfo {author} {\bibfnamefont {G.~E.}\ \bibnamefont {Scuseria}},\ }\href
  {\doibase 10.1063/1.2404663} {\bibfield  {journal} {\bibinfo  {journal} {The
  Journal of Chemical Physics}\ }\textbf {\bibinfo {volume} {125}},\ \bibinfo
  {pages} {224106} (\bibinfo {year} {2006})}\BibitemShut {NoStop}%
\bibitem [{\citenamefont {Togo}\ and\ \citenamefont {Tanaka}(2015)}]{Togo2015}%
  \BibitemOpen
  \bibfield  {author} {\bibinfo {author} {\bibfnamefont {A.}~\bibnamefont
  {Togo}}\ and\ \bibinfo {author} {\bibfnamefont {I.}~\bibnamefont {Tanaka}},\
  }\href {\doibase 10.1016/j.scriptamat.2015.07.021} {\bibfield  {journal}
  {\bibinfo  {journal} {Scripta Materialia}\ }\textbf {\bibinfo {volume}
  {108}},\ \bibinfo {pages} {1} (\bibinfo {year} {2015})}\BibitemShut {NoStop}%
\bibitem [{\citenamefont {Martyna}\ \emph {et~al.}(1992)\citenamefont
  {Martyna}, \citenamefont {Klein},\ and\ \citenamefont
  {Tuckerman}}]{Martyna1992}%
  \BibitemOpen
  \bibfield  {author} {\bibinfo {author} {\bibfnamefont {G.~J.}\ \bibnamefont
  {Martyna}}, \bibinfo {author} {\bibfnamefont {M.~L.}\ \bibnamefont {Klein}},
  \ and\ \bibinfo {author} {\bibfnamefont {M.}~\bibnamefont {Tuckerman}},\
  }\href {\doibase 10.1063/1.463940} {\bibfield  {journal} {\bibinfo  {journal}
  {The Journal of Chemical Physics}\ }\textbf {\bibinfo {volume} {97}},\
  \bibinfo {pages} {2635} (\bibinfo {year} {1992})}\BibitemShut {NoStop}%
\bibitem [{\citenamefont {Zhang}\ \emph {et~al.}(2021)\citenamefont {Zhang},
  \citenamefont {Wang}, \citenamefont {Guo}, \citenamefont {Li},\ and\
  \citenamefont {Wang}}]{Zhang2021}%
  \BibitemOpen
  \bibfield  {author} {\bibinfo {author} {\bibfnamefont {Y.}~\bibnamefont
  {Zhang}}, \bibinfo {author} {\bibfnamefont {B.}~\bibnamefont {Wang}},
  \bibinfo {author} {\bibfnamefont {Y.}~\bibnamefont {Guo}}, \bibinfo {author}
  {\bibfnamefont {Q.}~\bibnamefont {Li}}, \ and\ \bibinfo {author}
  {\bibfnamefont {J.}~\bibnamefont {Wang}},\ }\href {\doibase
  10.1016/j.commatsci.2021.110638} {\bibfield  {journal} {\bibinfo  {journal}
  {Computational Materials Science}\ }\textbf {\bibinfo {volume} {197}},\
  \bibinfo {pages} {110638} (\bibinfo {year} {2021})}\BibitemShut {NoStop}%
\bibitem [{\citenamefont {Sauvage}\ \emph {et~al.}(1982)\citenamefont
  {Sauvage}, \citenamefont {{De Backer}},\ and\ \citenamefont
  {Stymne}}]{Sauvage1982}%
  \BibitemOpen
  \bibfield  {author} {\bibinfo {author} {\bibfnamefont {F.~X.}\ \bibnamefont
  {Sauvage}}, \bibinfo {author} {\bibfnamefont {M.~G.}\ \bibnamefont {{De
  Backer}}}, \ and\ \bibinfo {author} {\bibfnamefont {B.}~\bibnamefont
  {Stymne}},\ }\href {\doibase 10.1016/0584-8539(82)80071-8} {\bibfield
  {journal} {\bibinfo  {journal} {Spectrochimica Acta Part A: Molecular
  Spectroscopy}\ }\textbf {\bibinfo {volume} {38}},\ \bibinfo {pages} {803}
  (\bibinfo {year} {1982})}\BibitemShut {NoStop}%
\bibitem [{\citenamefont {Wang}\ \emph {et~al.}(2021)\citenamefont {Wang},
  \citenamefont {Dong},\ and\ \citenamefont {Feng}}]{Wang2021}%
  \BibitemOpen
  \bibfield  {author} {\bibinfo {author} {\bibfnamefont {M.}~\bibnamefont
  {Wang}}, \bibinfo {author} {\bibfnamefont {R.}~\bibnamefont {Dong}}, \ and\
  \bibinfo {author} {\bibfnamefont {X.}~\bibnamefont {Feng}},\ }\href {\doibase
  10.1039/d0cs01160f} {\bibfield  {journal} {\bibinfo  {journal} {Chemical
  Society Reviews}\ }\textbf {\bibinfo {volume} {50}},\ \bibinfo {pages} {2764}
  (\bibinfo {year} {2021})}\BibitemShut {NoStop}%
\bibitem [{\citenamefont {Andrew}\ \emph {et~al.}(2012)\citenamefont {Andrew},
  \citenamefont {Mapasha}, \citenamefont {Ukpong},\ and\ \citenamefont
  {Chetty}}]{Andrew2012}%
  \BibitemOpen
  \bibfield  {author} {\bibinfo {author} {\bibfnamefont {R.~C.}\ \bibnamefont
  {Andrew}}, \bibinfo {author} {\bibfnamefont {R.~E.}\ \bibnamefont {Mapasha}},
  \bibinfo {author} {\bibfnamefont {A.~M.}\ \bibnamefont {Ukpong}}, \ and\
  \bibinfo {author} {\bibfnamefont {N.}~\bibnamefont {Chetty}},\ }\href
  {\doibase 10.1103/PhysRevB.85.125428} {\bibfield  {journal} {\bibinfo
  {journal} {Physical Review B}\ }\textbf {\bibinfo {volume} {85}},\ \bibinfo
  {pages} {125428} (\bibinfo {year} {2012})}\BibitemShut {NoStop}%
\bibitem [{\citenamefont {Born}\ and\ \citenamefont
  {Huang}(1954)}]{Born:224197}%
  \BibitemOpen
  \bibfield  {author} {\bibinfo {author} {\bibfnamefont {M.}~\bibnamefont
  {Born}}\ and\ \bibinfo {author} {\bibfnamefont {K.}~\bibnamefont {Huang}},\
  }\href {https://cds.cern.ch/record/224197} {\emph {\bibinfo {title}
  {{Dynamical theory of crystal lattices}}}},\ Oxford classic texts in the
  physical sciences\ (\bibinfo  {publisher} {Clarendon Press},\ \bibinfo
  {address} {Oxford},\ \bibinfo {year} {1954})\BibitemShut {NoStop}%
\bibitem [{\citenamefont {Liu}\ \emph {et~al.}(2017)\citenamefont {Liu},
  \citenamefont {Zhu}, \citenamefont {Ye},\ and\ \citenamefont
  {Yan}}]{Liu2017}%
  \BibitemOpen
  \bibfield  {author} {\bibinfo {author} {\bibfnamefont {C.-S.}\ \bibnamefont
  {Liu}}, \bibinfo {author} {\bibfnamefont {H.-H.}\ \bibnamefont {Zhu}},
  \bibinfo {author} {\bibfnamefont {X.-J.}\ \bibnamefont {Ye}}, \ and\ \bibinfo
  {author} {\bibfnamefont {X.-H.}\ \bibnamefont {Yan}},\ }\href {\doibase
  10.1039/C7NR00762K} {\bibfield  {journal} {\bibinfo  {journal} {Nanoscale}\
  }\textbf {\bibinfo {volume} {9}},\ \bibinfo {pages} {5854} (\bibinfo {year}
  {2017})}\BibitemShut {NoStop}%
\bibitem [{\citenamefont {Ettmayer}\ \emph {et~al.}(1978)\citenamefont
  {Ettmayer}, \citenamefont {Schebesta}, \citenamefont {Vendl},\ and\
  \citenamefont {Kieffer}}]{Ettmayer1978}%
  \BibitemOpen
  \bibfield  {author} {\bibinfo {author} {\bibfnamefont {P.}~\bibnamefont
  {Ettmayer}}, \bibinfo {author} {\bibfnamefont {W.}~\bibnamefont {Schebesta}},
  \bibinfo {author} {\bibfnamefont {A.}~\bibnamefont {Vendl}}, \ and\ \bibinfo
  {author} {\bibfnamefont {R.}~\bibnamefont {Kieffer}},\ }\href {\doibase
  10.1007/BF00907315} {\bibfield  {journal} {\bibinfo  {journal} {Monatshefte
  f{\"{u}}r Chemie - Chemical Monthly}\ }\textbf {\bibinfo {volume} {109}},\
  \bibinfo {pages} {929} (\bibinfo {year} {1978})}\BibitemShut {NoStop}%
\bibitem [{\citenamefont {Niwa}\ \emph {et~al.}(2019)\citenamefont {Niwa},
  \citenamefont {Yamamoto}, \citenamefont {Sasaki},\ and\ \citenamefont
  {Hasegawa}}]{Niwa2019}%
  \BibitemOpen
  \bibfield  {author} {\bibinfo {author} {\bibfnamefont {K.}~\bibnamefont
  {Niwa}}, \bibinfo {author} {\bibfnamefont {T.}~\bibnamefont {Yamamoto}},
  \bibinfo {author} {\bibfnamefont {T.}~\bibnamefont {Sasaki}}, \ and\ \bibinfo
  {author} {\bibfnamefont {M.}~\bibnamefont {Hasegawa}},\ }\href {\doibase
  10.1103/PhysRevMaterials.3.053601} {\bibfield  {journal} {\bibinfo  {journal}
  {Physical Review Materials}\ }\textbf {\bibinfo {volume} {3}},\ \bibinfo
  {pages} {053601} (\bibinfo {year} {2019})}\BibitemShut {NoStop}%
\bibitem [{\citenamefont {Reckeweg}\ \emph {et~al.}(2003)\citenamefont
  {Reckeweg}, \citenamefont {Lind}, \citenamefont {Simon},\ and\ \citenamefont
  {DiSalvo}}]{Reckeweg2003}%
  \BibitemOpen
  \bibfield  {author} {\bibinfo {author} {\bibfnamefont {O.}~\bibnamefont
  {Reckeweg}}, \bibinfo {author} {\bibfnamefont {C.}~\bibnamefont {Lind}},
  \bibinfo {author} {\bibfnamefont {A.}~\bibnamefont {Simon}}, \ and\ \bibinfo
  {author} {\bibfnamefont {F.~J.}\ \bibnamefont {DiSalvo}},\ }\href {\doibase
  10.1515/znb-2003-0124} {\bibfield  {journal} {\bibinfo  {journal}
  {Zeitschrift f{\"{u}}r Naturforschung B}\ }\textbf {\bibinfo {volume} {58}},\
  \bibinfo {pages} {159} (\bibinfo {year} {2003})}\BibitemShut {NoStop}%
\bibitem [{\citenamefont {BRADLEY}\ and\ \citenamefont
  {OLLARD}(1926)}]{BRADLEY1926}%
  \BibitemOpen
  \bibfield  {author} {\bibinfo {author} {\bibfnamefont {A.~J.}\ \bibnamefont
  {BRADLEY}}\ and\ \bibinfo {author} {\bibfnamefont {E.~F.}\ \bibnamefont
  {OLLARD}},\ }\href {\doibase 10.1038/117122b0} {\bibfield  {journal}
  {\bibinfo  {journal} {Nature}\ }\textbf {\bibinfo {volume} {117}},\ \bibinfo
  {pages} {122} (\bibinfo {year} {1926})}\BibitemShut {NoStop}%
\bibitem [{\citenamefont {Martin}\ and\ \citenamefont
  {Moore}(1959)}]{Martin1959}%
  \BibitemOpen
  \bibfield  {author} {\bibinfo {author} {\bibfnamefont {A.}~\bibnamefont
  {Martin}}\ and\ \bibinfo {author} {\bibfnamefont {A.}~\bibnamefont {Moore}},\
  }\href {\doibase 10.1016/0022-5088(59)90045-1} {\bibfield  {journal}
  {\bibinfo  {journal} {Journal of the Less Common Metals}\ }\textbf {\bibinfo
  {volume} {1}},\ \bibinfo {pages} {85} (\bibinfo {year} {1959})}\BibitemShut
  {NoStop}%
\bibitem [{\citenamefont {Miyahara}\ \emph {et~al.}(2007)\citenamefont
  {Miyahara}, \citenamefont {Kusuta},\ and\ \citenamefont
  {Furukawa}}]{Miyahara2007}%
  \BibitemOpen
  \bibfield  {author} {\bibinfo {author} {\bibfnamefont {S.}~\bibnamefont
  {Miyahara}}, \bibinfo {author} {\bibfnamefont {S.}~\bibnamefont {Kusuta}}, \
  and\ \bibinfo {author} {\bibfnamefont {N.}~\bibnamefont {Furukawa}},\ }\href
  {\doibase 10.1016/j.physc.2007.03.393} {\bibfield  {journal} {\bibinfo
  {journal} {Physica C: Superconductivity}\ }\textbf {\bibinfo {volume}
  {460-462}},\ \bibinfo {pages} {1145} (\bibinfo {year} {2007})}\BibitemShut
  {NoStop}%
\bibitem [{\citenamefont {Zhang}\ \emph {et~al.}(2010)\citenamefont {Zhang},
  \citenamefont {Hung},\ and\ \citenamefont {Wu}}]{Zhang2010}%
  \BibitemOpen
  \bibfield  {author} {\bibinfo {author} {\bibfnamefont {S.}~\bibnamefont
  {Zhang}}, \bibinfo {author} {\bibfnamefont {H.-h.}\ \bibnamefont {Hung}}, \
  and\ \bibinfo {author} {\bibfnamefont {C.}~\bibnamefont {Wu}},\ }\href
  {\doibase 10.1103/PhysRevA.82.053618} {\bibfield  {journal} {\bibinfo
  {journal} {Physical Review A}\ }\textbf {\bibinfo {volume} {82}},\ \bibinfo
  {pages} {053618} (\bibinfo {year} {2010})}\BibitemShut {NoStop}%
\bibitem [{\citenamefont {Wu}\ \emph {et~al.}(2007)\citenamefont {Wu},
  \citenamefont {Bergman}, \citenamefont {Balents},\ and\ \citenamefont {{Das
  Sarma}}}]{Wu2007}%
  \BibitemOpen
  \bibfield  {author} {\bibinfo {author} {\bibfnamefont {C.}~\bibnamefont
  {Wu}}, \bibinfo {author} {\bibfnamefont {D.}~\bibnamefont {Bergman}},
  \bibinfo {author} {\bibfnamefont {L.}~\bibnamefont {Balents}}, \ and\
  \bibinfo {author} {\bibfnamefont {S.}~\bibnamefont {{Das Sarma}}},\ }\href
  {\doibase 10.1103/PhysRevLett.99.070401} {\bibfield  {journal} {\bibinfo
  {journal} {Physical Review Letters}\ }\textbf {\bibinfo {volume} {99}},\
  \bibinfo {pages} {070401} (\bibinfo {year} {2007})}\BibitemShut {NoStop}%
\bibitem [{\citenamefont {Neupert}\ \emph {et~al.}(2011)\citenamefont
  {Neupert}, \citenamefont {Santos}, \citenamefont {Chamon},\ and\
  \citenamefont {Mudry}}]{Neupert2011}%
  \BibitemOpen
  \bibfield  {author} {\bibinfo {author} {\bibfnamefont {T.}~\bibnamefont
  {Neupert}}, \bibinfo {author} {\bibfnamefont {L.}~\bibnamefont {Santos}},
  \bibinfo {author} {\bibfnamefont {C.}~\bibnamefont {Chamon}}, \ and\ \bibinfo
  {author} {\bibfnamefont {C.}~\bibnamefont {Mudry}},\ }\href {\doibase
  10.1103/PhysRevLett.106.236804} {\bibfield  {journal} {\bibinfo  {journal}
  {Physical Review Letters}\ }\textbf {\bibinfo {volume} {106}},\ \bibinfo
  {pages} {236804} (\bibinfo {year} {2011})}\BibitemShut {NoStop}%
\bibitem [{\citenamefont {Ma}\ \emph {et~al.}(2009)\citenamefont {Ma},
  \citenamefont {Ji}, \citenamefont {Hu}, \citenamefont {Lu},\ and\
  \citenamefont {Xiang}}]{Ma2009}%
  \BibitemOpen
  \bibfield  {author} {\bibinfo {author} {\bibfnamefont {F.}~\bibnamefont
  {Ma}}, \bibinfo {author} {\bibfnamefont {W.}~\bibnamefont {Ji}}, \bibinfo
  {author} {\bibfnamefont {J.}~\bibnamefont {Hu}}, \bibinfo {author}
  {\bibfnamefont {Z.-Y.}\ \bibnamefont {Lu}}, \ and\ \bibinfo {author}
  {\bibfnamefont {T.}~\bibnamefont {Xiang}},\ }\href {\doibase
  10.1103/PhysRevLett.102.177003} {\bibfield  {journal} {\bibinfo  {journal}
  {Physical Review Letters}\ }\textbf {\bibinfo {volume} {102}},\ \bibinfo
  {pages} {177003} (\bibinfo {year} {2009})}\BibitemShut {NoStop}%
\bibitem [{\citenamefont {Yu}\ and\ \citenamefont {Si}(2015)}]{Yu2015}%
  \BibitemOpen
  \bibfield  {author} {\bibinfo {author} {\bibfnamefont {R.}~\bibnamefont
  {Yu}}\ and\ \bibinfo {author} {\bibfnamefont {Q.}~\bibnamefont {Si}},\ }\href
  {\doibase 10.1103/PhysRevLett.115.116401} {\bibfield  {journal} {\bibinfo
  {journal} {Physical Review Letters}\ }\textbf {\bibinfo {volume} {115}},\
  \bibinfo {pages} {116401} (\bibinfo {year} {2015})},\ \Eprint
  {http://arxiv.org/abs/1501.05926} {arXiv:1501.05926} \BibitemShut {NoStop}%
\bibitem [{\citenamefont {Ma}\ \emph {et~al.}(2008)\citenamefont {Ma},
  \citenamefont {Lu},\ and\ \citenamefont {Xiang}}]{Ma2008}%
  \BibitemOpen
  \bibfield  {author} {\bibinfo {author} {\bibfnamefont {F.}~\bibnamefont
  {Ma}}, \bibinfo {author} {\bibfnamefont {Z.-Y.}\ \bibnamefont {Lu}}, \ and\
  \bibinfo {author} {\bibfnamefont {T.}~\bibnamefont {Xiang}},\ }\href
  {\doibase 10.1103/PhysRevB.78.224517} {\bibfield  {journal} {\bibinfo
  {journal} {Physical Review B}\ }\textbf {\bibinfo {volume} {78}},\ \bibinfo
  {pages} {224517} (\bibinfo {year} {2008})}\BibitemShut {NoStop}%
\bibitem [{\citenamefont {Onsager}(1944)}]{Onsager1944}%
  \BibitemOpen
  \bibfield  {author} {\bibinfo {author} {\bibfnamefont {L.}~\bibnamefont
  {Onsager}},\ }\href {\doibase 10.1103/PhysRev.65.117} {\bibfield  {journal}
  {\bibinfo  {journal} {Physical Review}\ }\textbf {\bibinfo {volume} {65}},\
  \bibinfo {pages} {117} (\bibinfo {year} {1944})}\BibitemShut {NoStop}%
\bibitem [{\citenamefont {Yang}\ \emph {et~al.}(2018)\citenamefont {Yang},
  \citenamefont {Gong}, \citenamefont {Liu},\ and\ \citenamefont
  {Lu}}]{Yang2018}%
  \BibitemOpen
  \bibfield  {author} {\bibinfo {author} {\bibfnamefont {H.-C.}\ \bibnamefont
  {Yang}}, \bibinfo {author} {\bibfnamefont {B.-C.}\ \bibnamefont {Gong}},
  \bibinfo {author} {\bibfnamefont {K.}~\bibnamefont {Liu}}, \ and\ \bibinfo
  {author} {\bibfnamefont {Z.-Y.}\ \bibnamefont {Lu}},\ }\href {\doibase
  10.1016/j.scib.2018.05.036} {\bibfield  {journal} {\bibinfo  {journal}
  {Science Bulletin}\ }\textbf {\bibinfo {volume} {63}},\ \bibinfo {pages}
  {887} (\bibinfo {year} {2018})}\BibitemShut {NoStop}%
\bibitem [{\citenamefont {Xie}\ \emph {et~al.}(2012)\citenamefont {Xie},
  \citenamefont {Chen}, \citenamefont {Qin}, \citenamefont {Zhu}, \citenamefont
  {Yang},\ and\ \citenamefont {Xiang}}]{Xie2012}%
  \BibitemOpen
  \bibfield  {author} {\bibinfo {author} {\bibfnamefont {Z.~Y.}\ \bibnamefont
  {Xie}}, \bibinfo {author} {\bibfnamefont {J.}~\bibnamefont {Chen}}, \bibinfo
  {author} {\bibfnamefont {M.~P.}\ \bibnamefont {Qin}}, \bibinfo {author}
  {\bibfnamefont {J.~W.}\ \bibnamefont {Zhu}}, \bibinfo {author} {\bibfnamefont
  {L.~P.}\ \bibnamefont {Yang}}, \ and\ \bibinfo {author} {\bibfnamefont
  {T.}~\bibnamefont {Xiang}},\ }\href {\doibase 10.1103/PhysRevB.86.045139}
  {\bibfield  {journal} {\bibinfo  {journal} {Physical Review B}\ }\textbf
  {\bibinfo {volume} {86}},\ \bibinfo {pages} {045139} (\bibinfo {year}
  {2012})}\BibitemShut {NoStop}%
\bibitem [{\citenamefont {Torelli}\ and\ \citenamefont
  {Olsen}(2019)}]{Torelli2019}%
  \BibitemOpen
  \bibfield  {author} {\bibinfo {author} {\bibfnamefont {D.}~\bibnamefont
  {Torelli}}\ and\ \bibinfo {author} {\bibfnamefont {T.}~\bibnamefont
  {Olsen}},\ }\href {\doibase 10.1088/2053-1583/aaf06d} {\bibfield  {journal}
  {\bibinfo  {journal} {2D Materials}\ }\textbf {\bibinfo {volume} {6}},\
  \bibinfo {pages} {015028} (\bibinfo {year} {2019})}\BibitemShut {NoStop}%
\bibitem [{\citenamefont {Paier}\ \emph {et~al.}(2006)\citenamefont {Paier},
  \citenamefont {Marsman}, \citenamefont {Hummer}, \citenamefont {Kresse},
  \citenamefont {Gerber},\ and\ \citenamefont
  {{\'{A}}ngy{\'{a}}n}}]{Paier2006}%
  \BibitemOpen
  \bibfield  {author} {\bibinfo {author} {\bibfnamefont {J.}~\bibnamefont
  {Paier}}, \bibinfo {author} {\bibfnamefont {M.}~\bibnamefont {Marsman}},
  \bibinfo {author} {\bibfnamefont {K.}~\bibnamefont {Hummer}}, \bibinfo
  {author} {\bibfnamefont {G.}~\bibnamefont {Kresse}}, \bibinfo {author}
  {\bibfnamefont {I.~C.}\ \bibnamefont {Gerber}}, \ and\ \bibinfo {author}
  {\bibfnamefont {J.~G.}\ \bibnamefont {{\'{A}}ngy{\'{a}}n}},\ }\href {\doibase
  10.1063/1.2187006} {\bibfield  {journal} {\bibinfo  {journal} {The Journal of
  Chemical Physics}\ }\textbf {\bibinfo {volume} {124}},\ \bibinfo {pages}
  {154709} (\bibinfo {year} {2006})}\BibitemShut {NoStop}%
\bibitem [{\citenamefont {Guo}\ \emph {et~al.}(2020)\citenamefont {Guo},
  \citenamefont {Wang}, \citenamefont {Zhang}, \citenamefont {Yuan},
  \citenamefont {Ma},\ and\ \citenamefont {Wang}}]{Guo2020}%
  \BibitemOpen
  \bibfield  {author} {\bibinfo {author} {\bibfnamefont {Y.}~\bibnamefont
  {Guo}}, \bibinfo {author} {\bibfnamefont {B.}~\bibnamefont {Wang}}, \bibinfo
  {author} {\bibfnamefont {X.}~\bibnamefont {Zhang}}, \bibinfo {author}
  {\bibfnamefont {S.}~\bibnamefont {Yuan}}, \bibinfo {author} {\bibfnamefont
  {L.}~\bibnamefont {Ma}}, \ and\ \bibinfo {author} {\bibfnamefont
  {J.}~\bibnamefont {Wang}},\ }\href {\doibase 10.1002/inf2.12096} {\bibfield
  {journal} {\bibinfo  {journal} {InfoMat}\ }\textbf {\bibinfo {volume} {2}},\
  \bibinfo {pages} {639} (\bibinfo {year} {2020})}\BibitemShut {NoStop}%
\end{thebibliography}%

\end{document}